\documentclass[aps,prd,onecolumn,showpacs,showkeys,amsmath,amssymb,superscriptaddress,floatfix,nofootinbib]{revtex4}
\usepackage{amsfonts}
\usepackage{amsmath}
\usepackage{graphicx}
\usepackage{subfig}
\usepackage{dcolumn}
\usepackage{natbib}
\usepackage{bm}
\usepackage{booktabs}
\usepackage{slashed}

\usepackage{ulem}
\usepackage[usenames,dvipsnames]{color}

\makeatletter
\newcommand{\figcaption}{\def\@captype{figure}\caption}
\newcommand{\tabcaption}{\def\@captype{table}\caption}

\newcommand{\Rmnum}[1]{\expandafter\@slowromancap\romannumeral #1@}
\def\hlinewd#1{%
  \noalign{\ifnum0=`}\fi\hrule \@height #1 \futurelet
   \reserved@a\@xhline}
\makeatother
\def\dab{\int^{\alpha_{max}}_{\alpha_{min}}d\alpha\int^{\beta_{max}}_{\beta_{min}}d\beta}
\def\qq{\langle\bar qq\rangle}
\def\GGa{\langle GG\rangle}
\def\GGb{\langle g_s^2GG\rangle}
\def\qGqa{\langle\bar qg_s\sigma\cdot Gq\rangle}
\def\qGqb{\langle\bar qGq\rangle}
\def\f(s){[(\alpha+\beta)m^2-\alpha\beta s]}
\def\non{\\ \nonumber}

\begin{document}
%
%
\title{Exotic open-flavor $bc\bar{q}\bar{q}$, $bc\bar{s}\bar{s}$ and $qc\bar{q}\bar{b}$, $sc\bar{s}\bar{b}$ tetraquark states}

\author{Wei Chen}
\email{wec053@mail.usask.ca} \affiliation{Department of Physics and
Engineering Physics, University of Saskatchewan, Saskatoon, Saskatchewan, S7N
5E2, Canada}
\author{T. G. Steele}
\email{tom.steele@usask.ca} \affiliation{Department of Physics and
Engineering Physics, University of Saskatchewan, Saskatoon, SK, S7N
5E2, Canada}
\author{Shi-Lin Zhu}
\email{zhusl@pku.edu.cn}
\affiliation{Department of Physics and State Key Laboratory of Nuclear Physics and Technology,\\
Peking University, Beijing 100871, China, and \\
Collaborative Innovation Center of Quantum Matter, Beijing 100871,
China}

\begin{abstract}
We study the exotic $bc\bar{q}\bar{q}$, $bc\bar{s}\bar{s}$ and
$qc\bar{q}\bar{b}$, $sc\bar{s}\bar{b}$ systems by constructing the
corresponding tetraquark currents with $J^P=0^+$ and $1^+$. After
investigating the two-point correlation functions and the spectral
densities, we perform QCD sum rule analysis and extract the masses
of these open-flavor tetraquark states. Our results indicate that
the masses of both the scalar and axial vector tetraquark states are
about $7.1-7.2$ GeV for the $bc\bar{q}\bar{q}$ system and
$7.2-7.3$ GeV for the $bc\bar{s}\bar{s}$ system. For the
$qc\bar{q}\bar{b}$ tetraquark states with $J^P=0^+$ and $1^+$, their
masses are extracted to be around $7.1$ GeV. The masses for the
scalar and axial vector $sc\bar{s}\bar{b}$ states are $7.1$ GeV and
$6.9-7.1$ GeV, respectively. The tetraquark states $qc\bar{q}\bar{b}$
and $sc\bar{s}\bar{b}$ lie below the thresholds of $D^{(\ast)}B^{(\ast)}$
and $D_s^{(\ast)}B_s^{(\ast)}$ respectively, but they can decay into
$B_c$ plus a light meson. However, the tetraquark states $bc\bar{q}\bar{q}$
and $bc\bar{s}\bar{s}$ lie below the $D^{(\ast)}\bar B^{(\ast)}$ and
$D_s^{(\ast)}\bar B_s^{(\ast)}$ thresholds, suggesting dominantly weak decay mechanisms.
\end{abstract}


 \maketitle
%
%
 \section{Introduction}\label{sec:intro}
In the conventional quark model a meson is composed of a pair of
quark and antiquark ($q\bar q$) and a baryon is composed of three
quarks ($qqq$)~\cite{2007-Klempt-p1-202,2012-Beringer-p10001-10001}.
However, quantum chromodynamics (QCD) allows more complicated hadron
configurations. Hadrons with structures different from $q\bar q/qqq$
are sometimes called ``exotic" states. Although none of the exotic
states is now unambiguously identified, more and more unexpected
charmoniumlike and bottomoniumlike states have been observed in the
past several years. These resonances are considered as important
candidates of exotic hadrons, such as hadronic molecules, tetraquark
states, hybrids, etc.

The possible existence of the tetraquarks ($qq\bar q\bar q$) composed of a diquark and
an antidiquark was suggested by Jaffe in
1977~\cite{1977-Jaffe-p267-267,1977-Jaffe-p281-281}. The frequently
discussed candidates of tetraquark states are the light
scalars~\cite{1977-Jaffe-p267-267,1977-Jaffe-p281-281,2007-Chen-p94025-94025,2007-Chen-p369-372,2007-Zhang-p36004-36004}.
In the heavy quark sector, $qQ\bar q\bar Q$-type hidden-flavor
tetraquarks have been extensively studied to explain the underlying
structures of the recently observed $XYZ$ states in the relativistic
quark model~\cite{2006-Ebert-p214-219,2008-Ebert-p399-405}, QCD sum
rules~\cite{2007-Matheus-p14005-14005,2009-Bracco-p240-244,2010-Chen-p105018-105018,2011-Chen-p34010-34010,2013-Du-p33104-33104}
and via bound diquark clusters~\cite{2005-Maiani-p31502-31502,2007-Maiani-p182003-182003,2008-Maiani-p73004-73004,2013-Kleiv-p125018-125018}.
The existence and stability of doubly charmed/bottomed
$QQ\bar q\bar q$ tetraquark states have been also studied in the MIT
bag model~\cite{1988-Carlson-p744-744}, chiral quark model~\cite{2008-Zhang-p437-440, 1997-Pepin-p119-123}, constituent
quark model~\cite{2006-Vijande-p54018-54018,2009-Vijande-p74010-74010,1998-Brink-p6778-6787,
1993-Silvestre-Brac-p457-470,1986-Zouzou-p457-457}, relativistic quark model~\cite{2007-Ebert-p114015-114015}, chiral
perturbation theory~\cite{1993-Manohar-p17-33}, QCD sum
rules~\cite{2007-Cui-p7-13,2007-Navarra-p166-172,2013-Du-p14003-14003,2011-Wang-p1049-1058} and some other
methods~\cite{1973-Lipkin-p267-271,1982-Ader-p2370-2370,1986-Lipkin-p242-242,1991-Richard-p254-257,1994-Bander-p5478-5480,1996-Moinester-p349-362,
2003-Gelman-p296-304,2011-Carames-p291-295}.

Recently, there have been efforts to understand the open-flavor
(i.e., exotic)
tetraquark states $bc\bar q\bar
q$~\cite{1993-Silvestre-Brac-p457-470,1986-Zouzou-p457-457,2013-Feng-p-}
and molecular states $\bar qc\bar b
q$~\cite{2009-Zhang-p56004-56004,2012-Sun-p94008-94008,2012-Albuquerque-p492-498}.
The authors of Ref.~\cite{2013-Feng-p-} noticed that the tetraquark
states $bc\bar q\bar q$ lie below the thresholds of $B^-D^+$ and
$\bar B^0D^0$ by solving the Bethe-Salpeter equations. In
Refs.~\cite{2012-Sun-p94008-94008,2012-Albuquerque-p492-498}, the
authors indicated that there may exist loosely bound $B_c$-like
molecular states. In this paper, we will study the open-flavor
$bc\bar{q}\bar{q}$, $bc\bar{s}\bar{s}$ and $qc\bar{q}\bar{b}$,
$sc\bar{s}\bar{b}$ tetraquark states in QCD sum rules. We construct
the corresponding tetraquark currents with $J^P=0^+, 1^+$ by using
$S$-wave diquark fields. With these interpolating operators, we
calculate the two-point correlation functions and extract the masses
of these possible tetraquark states.

This paper is organized as follows. In Sec.~\ref{sec:current}, we
construct all the scalar and axial vector $bc\bar{q}\bar{q},
bc\bar{s}\bar{s}$ and $qc\bar{q}\bar{b}$, $sc\bar{s}\bar{b}$ types
of tetraquark currents with $S$-wave diquark fields and the
corresponding antidiquark fields. In Sec.~\ref{sec:QSR}, we
calculate the two-point correlation functions and the spectral
densities using these interpolating tetraquark currents. The
expressions for the spectral densities are listed in the
Appendix. We perform QCD sum rule analysis for these
tetraquark systems and extract their masses in Sec.~\ref{sec:NA}. We
also construct mixed interpolating currents to study
mixing effects. In the final
section, we summarize our results and discuss the possible decay
properties of these tetraquark states.

\section{tetraquark interpolating currents}\label{sec:current}
In this section, we construct the $bc\bar{q}\bar{q}$ and
$qc\bar{q}\bar{b}$ types of tetraquark interpolating currents using
diquark and antidiquark fields. In general, one can use the
diquark fields $q^T_aCq_b$, $q^T_a C\gamma_5q_b$, $q^T_aC\gamma_\mu
q_b$, $q^T_aC\gamma_\mu\gamma_5q_b$, $q^T_a C\sigma_{\mu\nu}q_b$,
$q^T_aC\sigma_{\mu\nu}\gamma_5q_b$ and the corresponding
antidiquark fields to compose all possible combinations of
$bc\bar{q}\bar{q}$ and $qc\bar{q}\bar{b}$ tetraquark operators, as
done in Ref.~\cite{2013-Du-p14003-14003} for the
doubly charmed/bottomed tetraquark states and
Refs.~\cite{2010-Chen-p105018-105018,2011-Chen-p34010-34010} for 
the charmoniumlike and bottomoniumlike tetraquark states. In
Ref.~\cite{2013-Du-p14003-14003}, the tetraquark currents which
contain $P$-wave diquark or antidiquark operators can result in
higher hadron masses than those containing only $S$-wave operators.
They correspond to the orbitally excited states while the latter
operators correspond to the ground hadron states. In order to study 
the lowest lying tetraquark states, we use only the $S$-wave diquark 
fields $q_a^TC\gamma_5q_b$, $q_a^TC\gamma_{\mu}q_b$ and the
corresponding antidiquark fields to compose the tetraquark currents
with quantum numbers $J^P=0^+, 1^+$. The $P$-wave diquark fields 
will not be considered in this paper.

For the $bc\bar q\bar q$ system, the tetraquark interpolating
currents with $J^P=0^+$ are
\begin{equation}
\begin{split}
J_1&=b^T_aC\gamma_5c_b(\bar{q}_a\gamma_5C\bar{q}^T_b+\bar{q}_b\gamma_5C\bar{q}^T_a),\\
J_2&=b^T_aC\gamma_\mu c_b(\bar{q}_a\gamma^\mu
C\bar{q}^T_b+\bar{q}_b\gamma^\mu C\bar{q}^T_a),
\\
J_3&=b^T_aC\gamma_5c_b(\bar{q}_a\gamma_5C\bar{q}^T_b-\bar{q}_b\gamma_5C\bar{q}^T_a),\\
J_4&=b^T_aC\gamma_\mu c_b(\bar{q}_a\gamma^\mu
C\bar{q}^T_b-\bar{q}_b\gamma^\mu C\bar{q}^T_a), \label{current1}
\end{split}
\end{equation}
in which ``$+$" denotes the symmetric color structure
$[\mathbf{6_c}]_{bc} \otimes [\mathbf{ \bar 6_c}]_{\bar{q}\bar q}$
and ``$-$" denotes the antisymmetric color structure $[\mathbf{\bar
3_c}]_{bc} \otimes [\mathbf{3_c}]_{\bar{q}\bar q}$. The tetraquark
interpolating currents with $J^P=1^+$ are
\begin{equation}
\begin{split}
J_{1\mu}&=b^T_aC\gamma_5c_b(\bar{q}_a\gamma_\mu C\bar{q}_b^T+\bar{q}_b\gamma_\mu C\bar{q}_a^T),\\
J_{2\mu}&=b^T_aC\gamma_\mu c_b(\bar{q}_a\gamma_5C\bar{q}_b^T+\bar{q}_b\gamma_5C\bar{q}^T_a),\\
J_{3\mu}&=b^T_aC\gamma_5c_b(\bar{q}_a\gamma_\mu C\bar{q}_b^T-\bar{q}_b\gamma_\mu C\bar{q}_a^T),\\
J_{4\mu}&=b^T_aC\gamma_\mu
c_b(\bar{q}_a\gamma_5C\bar{q}_b^T-\bar{q}_b\gamma_5C\bar{q}^T_a),
\label{current2}
\end{split}
\end{equation}
where  ``$+$" again denotes the symmetric color structure
$[\mathbf{6_c}]_{bc} \otimes [\mathbf{ \bar 6_c}]_{\bar{q}\bar q}$
and ``$-$" denotes the antisymmetric color structure $[\mathbf{\bar
3_c}]_{bc} \otimes [\mathbf{3_c}]_{\bar{q}\bar q}$.

Similarly, for the $cq\bar b\bar q$ system, the tetraquark interpolating
currents with $J^P=0^+$ are
\begin{equation}
\begin{split}
J_1&=q^T_aC\gamma_5c_b(\bar{q}_a\gamma_5C\bar{b}^T_b+\bar{q}_b\gamma_5C\bar{b}^T_a),\\
J_2&=q^T_aC\gamma_\mu c_b(\bar{q}_a\gamma^\mu
C\bar{b}^T_b+\bar{q}_b\gamma^\mu C\bar{b}^T_a),
\\
J_3&=q^T_aC\gamma_5c_b(\bar{q}_a\gamma_5C\bar{b}^T_b-\bar{q}_b\gamma_5C\bar{b}^T_a),\\
J_4&=q^T_aC\gamma_\mu c_b(\bar{q}_a\gamma^\mu
C\bar{b}^T_b-\bar{q}_b\gamma^\mu C\bar{b}^T_a), \label{current3}
\end{split}
\end{equation}
in which ``$+$" denotes the symmetric color structure
$[\mathbf{6_c}]_{qc} \otimes [\mathbf{ \bar 6_c}]_{\bar{q}\bar b}$
and ``$-$" denotes the antisymmetric color structure $[\mathbf{\bar
3_c}]_{qc} \otimes [\mathbf{3_c}]_{\bar{q}\bar b}$. The tetraquark
interpolating currents with $J^P=1^+$ are
\begin{equation}
\begin{split}
J_{1\mu}&=q^T_aC\gamma_5c_b(\bar{q}_a\gamma_\mu C\bar{b}_b^T+\bar{q}_b\gamma_\mu C\bar{b}_a^T),\\
J_{2\mu}&=q^T_aC\gamma_\mu c_b(\bar{q}_a\gamma_5C\bar{b}_b^T+\bar{q}_b\gamma_5C\bar{b}^T_a),\\
J_{3\mu}&=q^T_aC\gamma_5c_b(\bar{q}_a\gamma_\mu C\bar{b}_b^T-\bar{q}_b\gamma_\mu C\bar{b}_a^T),\\
J_{4\mu}&=q^T_aC\gamma_\mu
c_b(\bar{q}_a\gamma_5C\bar{b}_b^T-\bar{q}_b\gamma_5C\bar{b}^T_a),
\label{current4}
\end{split}
\end{equation}
where ``$+$" again denotes the symmetric color structure
$[\mathbf{6_c}]_{qc} \otimes [\mathbf{ \bar 6_c}]_{\bar{q}\bar b}$
and ``$-$" denotes the antisymmetric color structure $[\mathbf{\bar
3_c}]_{qc} \otimes [\mathbf{3_c}]_{\bar{q}\bar b}$.

Replacing the light quark $q$ by the strange quark $s$ in
Eqs.~\eqref{current3} and \eqref{current4}, we can also obtain the
corresponding $cs\bar b\bar s$ tetraquark currents with the same
quantum numbers. However, the $bc\bar s\bar s$ system is different.
In this system, the flavor structure of $\bar s\bar s$ pair is
symmetric and thus its color structure is fixed at the same time.
The color structures for the diquark fields $s_a^TC\gamma_5s_b$ and
$s_a^TC\gamma_{\mu}s_b$ are symmetric $\mathbf{6_c}$ and
antisymmetric $\mathbf{\bar 3_c}$, respectively. As a result, only
$J_1$, $J_4$ in Eq.~\eqref{current1} and $J_{2\mu}$, $J_{3\mu}$ in
Eq.~\eqref{current2} survive in the $bc\bar s\bar s$ system and all
the other currents vanish.

\section{Two-Point Correlation Function and Spectral Density}\label{sec:QSR}
In the framework of QCD sum
rules~\cite{1979-Shifman-p385-447,1985-Reinders-p1-1,2000-Colangelo-p1495-1576},
we consider the two-point correlation functions
\begin{eqnarray}
\Pi(p^{2})&=& i\int d^4xe^{ip\cdot
x}\langle0|T[J(x)J^{\dag}(0)]|0\rangle, \label{equ:Pi1}
\\ \Pi_{\mu\nu}(p^{2})&=& i\int d^4xe^{ip\cdot x}\langle0|T[J_{\mu}(x)J_{\nu}^{\dag}(0)]|0\rangle, \label{equ:Pi2}
\end{eqnarray}
where $J(x)$ and $J_{\mu}(x)$ are the scalar and axial vector
currents shown in Eqs.~\eqref{current1}--\eqref{current4}. Since the
axial vector currents $J_{\mu}(x)$ are not conserved,  the two-point
correlation function $\Pi_{\mu\nu}(p^{2})$ has the following
structure
\begin{eqnarray}
\Pi_{\mu\nu}(p^{2})=\left(\frac{p_{\mu}p_{\nu}}{p^2}-g_{\mu\nu}\right)\Pi_1(p^2)+\frac{p_{\mu}p_{\nu}}{p^2}\Pi_0(p^2),
\end{eqnarray}
where $\Pi_1(p^2)$ and $\Pi_0(p^2)$ are the invariant functions
related to the spin-1 and spin-0 intermediate states, respectively.
In this paper, we focus on $\Pi_1(p^2)$ to study the axial vector
channels.

In QCD sum rules, the correlation functions in Eqs.~\eqref{equ:Pi1}
and \eqref{equ:Pi2} can be obtained at both the hadron level and
quark-gluon level. At the hadron level, we can describe the
correlation function via the dispersion relation
\begin{eqnarray}
\Pi(p^2)=(p^2)^N\int_{(m_c+m_b)^2}^{\infty}\frac{\rho(s)}{s^N(s-p^2-i\epsilon)}ds+\sum_{n=0}^{N-1}b_n(p^2)^n,
\label{dispersionrelation}
\end{eqnarray}
in which $b_n$ are the $N$ unknown subtraction constants which can be
removed by taking the Borel transform. To obtain the spectral function
$\rho(s)$, we write the imaginary part of $\Pi(p^2)$ as a sum over
$\delta$ functions by inserting intermediate hadronic states
$|n\rangle$ with the same quantum numbers as the interpolating
current $J(x)$,
\begin{eqnarray}
\nonumber
\rho(s)&\equiv&\frac{1}{\pi}\text{Im}\Pi(s)=\sum_n\delta(s-m_n^2)\langle0|J|n\rangle\langle
n|J^{\dagger}|0\rangle
\\&=&f_X^2\delta(s-m_X^2)+ \mbox{continuum},  \label{Phenrho}
\end{eqnarray}
where we adopt the pole plus continuum parametrization of the
hadronic spectral density and $m_X$ is the mass of the lowest lying
resonance $|X\rangle$. The scalar and axial vector interpolating
currents $J(x)$ and $J_{\mu}(x)$ can couple to the corresponding
hadronic states with the coupling parameters $f_X$,
\begin{eqnarray}
\langle0|J|X\rangle&=&f_X, \label{coupling parameter1}
\\
\langle0|J_{\mu}|X\rangle&=&f_X\epsilon_{\mu}, \label{coupling
parameter2}
\end{eqnarray}
where $\epsilon_{\mu}$ is the polarization vector ($\epsilon\cdot
p=0$).

The correlation function can also be evaluated at the quark-gluon
level via the operator product expansion (OPE) method. We calculate
the Wilson coefficients up to dimension eight at leading order in
$\alpha_s$. Utilizing the same technique as in
Refs.~\cite{2010-Chen-p105018-105018,2011-Chen-p34010-34010,2013-Du-p33104-33104,2013-Du-p14003-14003,2013-Chen-p-},
we adopt the coordinate expression for the light quark propagator
and the momentum space expression for the heavy quark propagator,
\begin{eqnarray}
\nonumber iS^{ab}_q(x) &=&  \frac{i\delta^{ab}}{2\pi^2x^4}\hat{x}
+\frac{i}{32\pi^2}\frac{\lambda^n_{ab}}{2}g_sG_{\mu\nu}^n\frac{1}
{x^2}(\sigma^{\mu\nu}\hat{x}+\hat{x}\sigma^{\mu\nu})-\frac{\delta^{ab}}{12}\langle\bar{q}q\rangle+
\frac{\delta^{ab}x^2}{192}\qGqa-\frac{m_q\delta^{ab}}{4\pi^2x^2}
\\&&+\frac{i\delta^{ab}m_q\langle\bar{q}q\rangle}{48}\hat{x}
-\frac{im_q\qGqa\delta^{ab}x^2\hat{x}}{1152},
\label{prop_light}
\\
iS^{ab}_Q(p) &=&
\frac{i\delta^{ab}}{\hat{p}-m_Q}+\frac{i}{4}g_s\frac{\lambda^n_{ab}}{2}G_{\mu\nu}^n
\frac{\sigma^{\mu\nu}(\hat{p}+m_Q)+(\hat{p}+m_Q)\sigma^{\mu\nu}}
{(p^2-m_Q^2)^2}+\frac{i\delta^{ab}}{12}\GGb
m_Q\frac{p^2+m_Q\hat{p}}{(p^2-m_Q^2)^4},
\end{eqnarray}
in which $q$ represents $u$, $d$ or $s$ quark and $Q$ represents $c$
or $b$ quark. The superscripts $a, b$ are color indices and $\hat
x=\gamma_{\mu}x^{\mu}$, $\hat p=\gamma_{\mu}p^{\mu}$. We keep the
terms proportional to $m_q$ to study the $sc\bar{s}\bar{b}$ and
$bc\bar{s}\bar{s}$ systems. In particular, the $m_s$ corrections 
are only important for the chiral-violating condensates; the $m_q$ 
corrections to the gluon condensate that would arise from an $m_q$ 
term in \eqref{prop_light}  are numerically small and are thus 
ignored (see Fig.~\ref{figOPE} below).

By equating the correlation functions at both the hadron level and
quark-gluon level, we can establish the sum rules for the hadron
parameters via quark-hadron duality. Using the spectral function
defined in Eq.~\eqref{Phenrho}, the Borel transform is performed on
the correlation function $\Pi(p^2)$ obtained at both levels to
remove the unknown constants in Eq.~\eqref{dispersionrelation},
improve the convergence of the OPE series and suppress the continuum
contributions
\begin{eqnarray}
\mathcal{L}_{k}\left(s_0,
M_B^2\right)=f_X^2m_X^{2k}e^{-m_X^2/M_B^2}=\int_{(m_c+m_b)^2}^{s_0}dse^{-s/M_B^2}\rho(s)s^k,
\label{sumrule}
\end{eqnarray}
where $s_0$ is the continuum threshold parameter and $M_B$ is the
Borel mass introduced by the Borel transform. These two parameters
are very important in QCD sum rule analysis and we  will discuss them
carefully in the next section. Then the mass of the lowest lying
hadron state can be extracted as
\begin{eqnarray}
m_X\left(s_0, M_B^2\right)=\sqrt{\frac{\mathcal{L}_{1}\left(s_0,
M_B^2\right)}{\mathcal{L}_{0}\left(s_0, M_B^2\right)}}, \label{mass}
\end{eqnarray}
which is a function of the continuum threshold $s_0$ and Borel mass
$M_B$. At the leading order in $\alpha_s$, the spectral densities
for all interpolating currents in 
Eqs.~\eqref{current1}--\eqref{current4} are evaluated and listed in
the Appendix up to  dimension eight condensates. For the
nonperturbative contributions, the quark condensate $\qq$, gluon
condensate $\GGa$, quark-gluon condensate mixed $\qGqa$, four quark
condensate and dimension eight condensate contribute to the correlation
functions and spectral densities.
Using the factorization hypothesis, the dimension six and eight
condensates are reduced to $\qq^2$ and $\qq\qGqa$ respectively.
The evaluation of the higher dimension condensate contributions is
technically difficult and the violation of the factorization hypothesis
becomes important~\cite{1992-Braaten-p581-612}. In this paper, we calculate
the correlation functions up to dimension eight.

\section{QCD Sum Rule Analysis}\label{sec:NA}
To perform the QCD sum rule analysis, we adopt the following values
of the quark masses and various
condensates~\cite{1985-Reinders-p1-1,2012-Beringer-p10001-10001,2012-Narison-p259-263,2010-Narison-p559-559,2007-Kuhn-p192-215}
in the chiral limit $(m_u=m_d=0)$:
\begin{eqnarray}
\nonumber && m_s(2\,\text{GeV})=(101^{+29}_{-21})\text{ MeV} \, , \non
&&m_c(\mu=m_c)=\overline m_c=(1.28\pm 0.02)~\mbox{GeV}   \, , \non
&&m_b(\mu=m_b)=\overline m_b=(4.17\pm 0.02)~\mbox{GeV}   \, , \non
&&\qq=-(0.23\pm0.03)^3\text{ GeV}^3 \, ,
\\   &&\qGqa=-M_0^2\qq\, ,
\non &&M_0^2=(0.8\pm0.2)\text{ GeV}^2 \, , \non &&\langle\bar
ss\rangle/\qq=0.8\pm0.1 \, , \non &&\GGb=(0.48\pm0.14)\text{GeV}^4
\, , \label{parameters}
\end{eqnarray}
in which the definition of the coupling constant $g_s$ has a minus
sign difference compared to that in Ref.~\cite{1985-Reinders-p1-1}.
The charm and bottom quark masses are the
running masses in the $\overline{\rm MS}$ scheme. Furthermore, we
take into account the scale dependence of these $\overline{\rm MS}$
masses at leading order:
\begin{eqnarray}
m_c(\mu)=\overline m_c\bigg(\frac{\alpha_s(\mu)}{\alpha_s(\overline
m_c)}\bigg)^{12/25},
\\m_b(\mu)=\overline m_b\bigg(\frac{\alpha_s(\mu)}{\alpha_s(\overline m_b)}\bigg)^{12/23}\, ,
\end{eqnarray}
where
\begin{eqnarray}
\alpha_s(\mu)&=&\frac{\alpha_s(M_{\tau})}{1+\frac{25\alpha_s(M_{\tau})}{12\pi}\log(\frac{\mu^2}{M_{\tau}^2})},
\quad \alpha_s(M_{\tau})=0.33, \label{alpha_cc}
\end{eqnarray}
is determined by evolution from the $\tau$ mass using Particle Data
Group values~\cite{2012-Beringer-p10001-10001}. For the
$bc\bar{q}\bar{q}$ and $qc\bar{q}\bar{b}$ tetraquark systems, we use
the renormalization scale $\mu=\frac{\overline m_c+\overline m_b}{2}=2.73$ GeV in
our sum rule analysis~\cite{2013-Chen-p-a}.

After performing the Borel transform, there are two important parameters
in the correlation function: the continuum threshold $s_0$ and the
Borel mass $M_B$. The stability of QCD sum rules requires a suitable
working region of these two parameters. In our analysis, we choose
the value of $s_0$ to minimize the variation of the extracted mass
$m_X$ with the Borel mass $M_B^2$. Using this value of $s_0$, we can
obtain a suitable Borel window by studying the convergence of the
OPE series and pole contribution. The requirement of the OPE
convergence determines a lower bound on $M_B^2$ while the constraint
of the pole contribution leads to its upper bound.

The pole contribution (PC) is defined as
\begin{eqnarray}
\text{PC}(s_0, M_B^2)=\frac{\mathcal{L}_{0}\left(s_0,
M_B^2\right)}{\mathcal{L}_{0}\left(\infty, M_B^2\right)},
\label{PC}
\end{eqnarray}
which is a function of the continuum threshold $s_0$ and the Borel
mass $M_B$. This definition comes from the sum rules established in
Eq.~\eqref{sumrule} and indicates the contribution of the lowest
lying resonance to the correlation function.

\subsection{$bc\bar q\bar q$ and $bc\bar s\bar s$ tetraquark systems}
We begin with the sum rule analysis of the $bc\bar q\bar q$ and
$bc\bar s\bar s$ tetraquark systems in this subsection. For all
currents in the $bc\bar q\bar q$ systems, the quark condensate $\qq$
and quark gluon mixed condensate $\qGqa$ terms in the correlation
functions are proportional to the light quark mass $m_q$. Both of
them vanish in chiral limit $m_q=0$ and represent a numerically
small contribution to the correlation functions because of this
chiral suppression. For these systems, the four quark condensate
$\qq^2$ is the dominant power correction to the correlation
function. We show the OPE convergence of the scalar $bc\bar q\bar q$
channel using the interpolating current $J_4$ in Fig.~\ref{figOPE}.
It indicates that the dimension eight condensate $\qq\qGqa$ is
the next in importance followed by the gluon condensate $\GGa$.
To ensure the convergence of the
OPE series, we require that the four quark condensate contribution
be less than one-fifth of the perturbative term, which results in a
lower bound on the Borel mass $M_B$. In Fig.~\ref{figOPE}, the OPE
convergence is very good in the region $M_B^2\geq6.1$ GeV$^2$. This
value is the lower bound on $M_B^2$ for $J_4$ scalar channel of
$bc\bar q\bar q$ system.
\begin{center}
\includegraphics[scale=0.73]{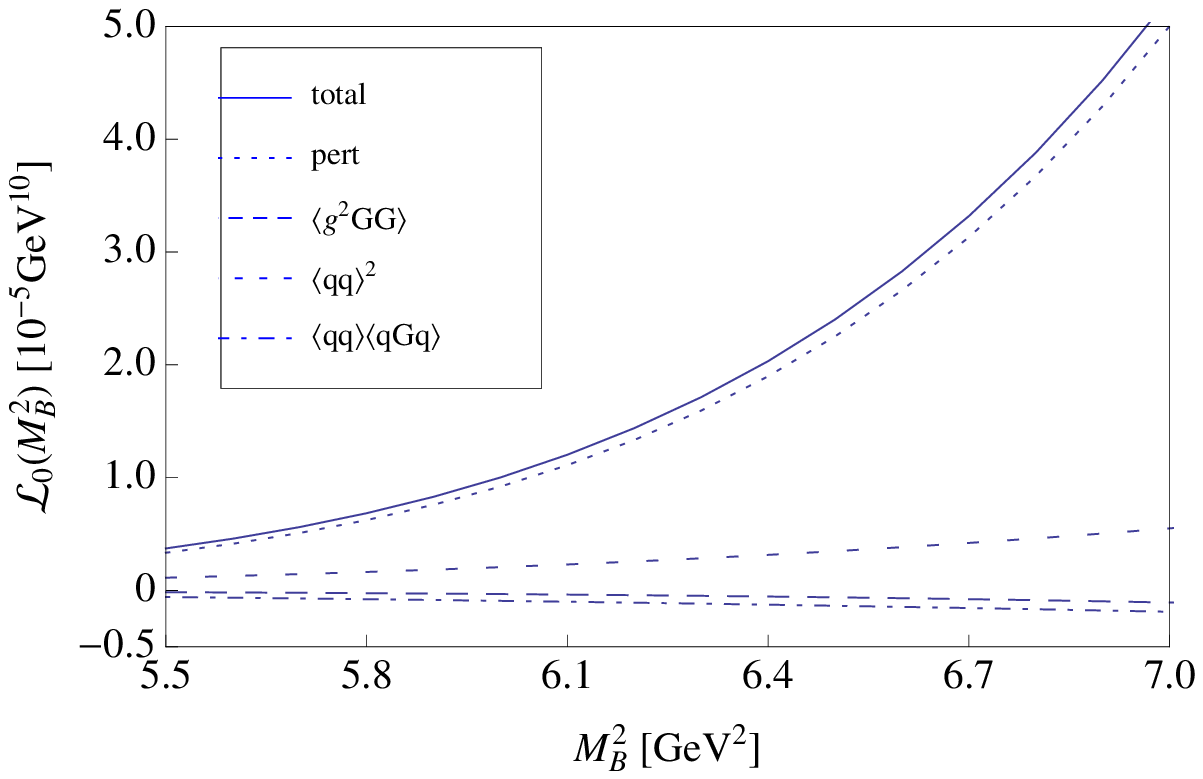}
\figcaption{OPE convergence for the current $J_4$ in the $J^P=0^+$
$bc\bar q\bar q$ system.} \label{figOPE}
\end{center}

\begin{center}
\begin{tabular}{lr}
\scalebox{0.65}{\includegraphics{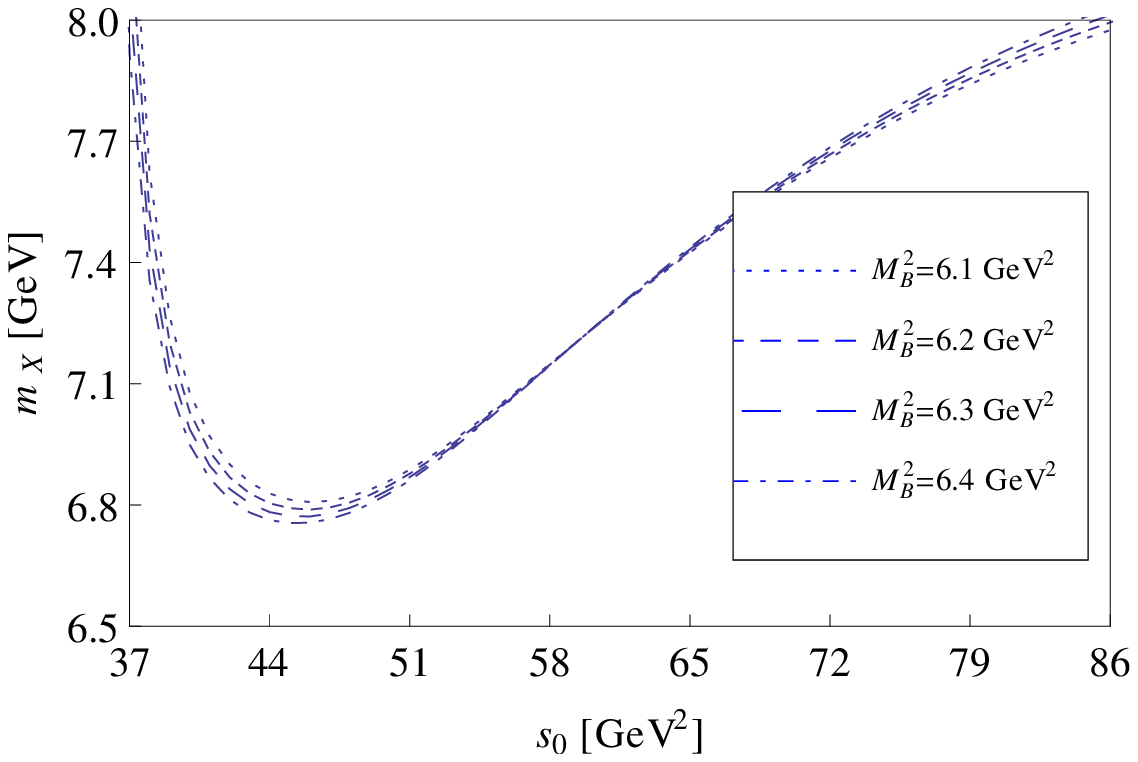}}&
\scalebox{0.65}{\includegraphics{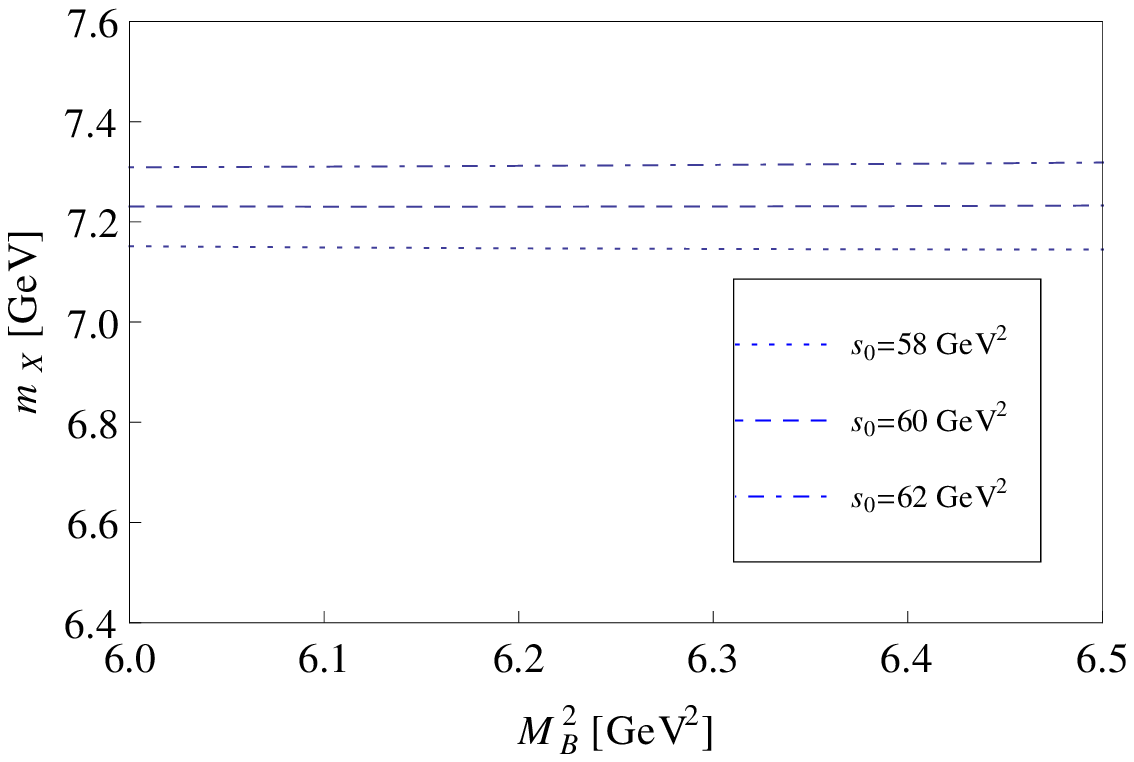}}
\end{tabular}
\figcaption{Variation of $m_X$ with $s_0$ and $M^2_B$ corresponding
to the current $J_4$ for the $0^{+}$ $bc\bar q\bar q$ system.}
\label{figbcqq0+4}
\end{center}

On the other hand, an upper bound on $M_B^2$ is obtained by studying
the pole contribution defined in Eq.~\eqref{PC}, which is also the
function of the continuum threshold $s_0$. To study the variation of
PC with $M_B$, one should determine the value of $s_0$ at first. An
optimized choice of $s_0$ is the value minimizing the variation of
the extracted hadron mass $m_X$ with the Borel parameter $M_B^2$. We
study this in the left portion of Fig.~\ref{figbcqq0+4} for the scalar
$bc\bar q\bar q$ channel with the current $J_4$. Varying the value
of $M_B^2$ from its lower bound $M_{min}^2=6.1$ GeV$^2$, these mass
curves with different value of $M_B^2$ intersect at $s_0=60$ GeV$^2$, 
which is the most suitable value under the above constraint. Utilizing 
this value of $s_0$, we require that PC be larger than $30\%$ to determine 
the upper bound on the Borel mass $M_B^2$. For the current $J_4$ in the 
scalar $bc\bar q\bar q$ channel, we obtain the upper bound $M_{max}^2=6.4$ 
GeV$^2$.

For the $J^P=0^+$ $bc\bar q\bar q$ systems, all currents $J_1, J_2,
J_3$ and $J_4$ have suitable working range of the Borel parameter
with the above criteria. Within these Borel windows, the mass sum
rules are very stable. In Fig.~\ref{figbcqq0+4}, we show the
variation of $m_X$ with the threshold value $s_0$ and Borel
parameter $M^2_B$ for the current $J_4$. We obtain the Borel
window $6.1$ GeV$^2$ $\leq M_B^2\leq 6.4$ GeV$^2$ with the continuum
threshold value $s_0=60$ GeV$^2$. In this region, we show the stable
mass sum rule in the right portion of Fig.~\ref{figbcqq0+4} and extract
the hadron mass 
\begin{eqnarray}
m_X=7.23\pm0.08\pm0.05\pm0.06~\text{GeV},
\end{eqnarray}
in which the errors come respectively from the continuum threshold $s_0$,
the heavy quark masses $m_c, m_b$ and the quark condensates $\qq, \qGqa$.
The errors from the Borel mass $M_B$ and the gluon condensate $\GGb$ are
negligible since the mass sum rules are very stable in the Borel window
(see Fig.~\ref{figbcqq0+4} and Fig.~\ref{figbcss0+4}) while, as mentioned 
above, the gluon condensate contribution to the correlation function is very small.

After performing the QCD sum rule analyses for all the interpolating
currents, we collect the Borel window, the threshold value, the
extracted mass and the pole contribution for the $J^P=0^+$ $bc\bar
q\bar q$ systems in Table~\ref{tablebcqq0+}. The results for the
$J^P=1^+$ $bc\bar q\bar q$ systems are listed in Table~\ref{tablebcqq1+}.
As mentioned above, the errors of mass predictions come from the
uncertainties in $s_0$, the heavy quark masses $m_c, m_b$ and QCD condensates
$\qq, \qGqa$ respectively.

\begin{center}
\begin{tabular}{lr}
\scalebox{0.65}{\includegraphics{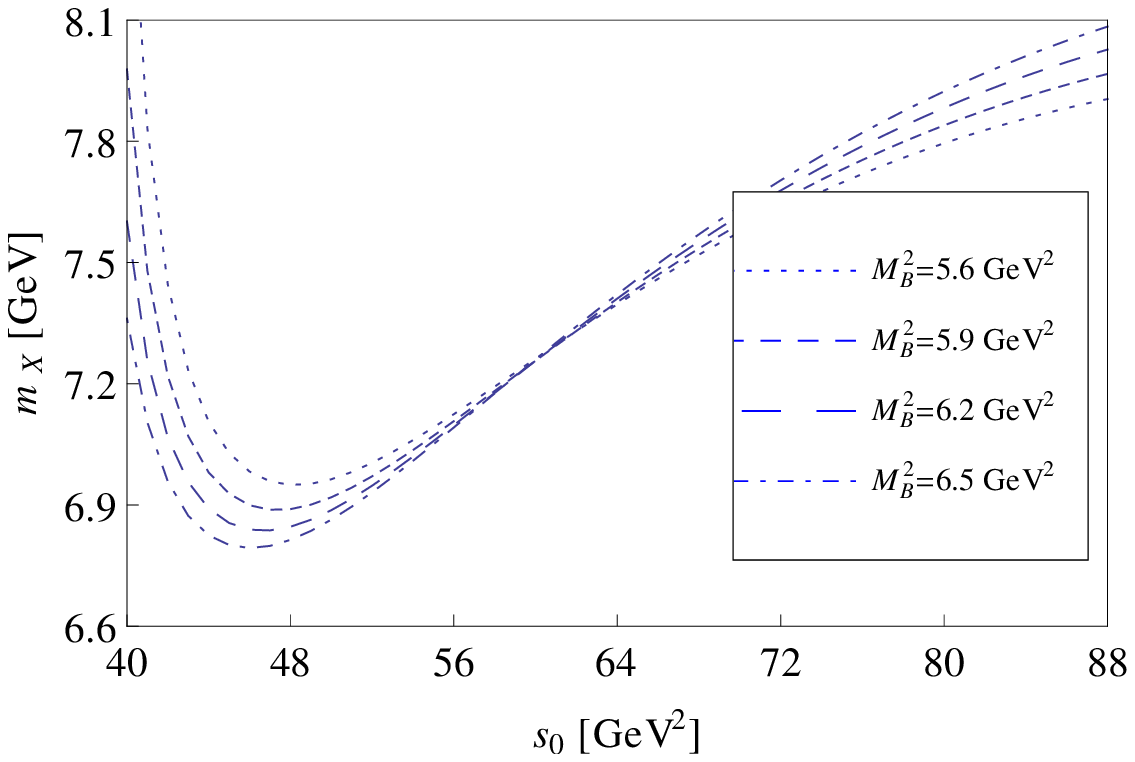}}&
\scalebox{0.65}{\includegraphics{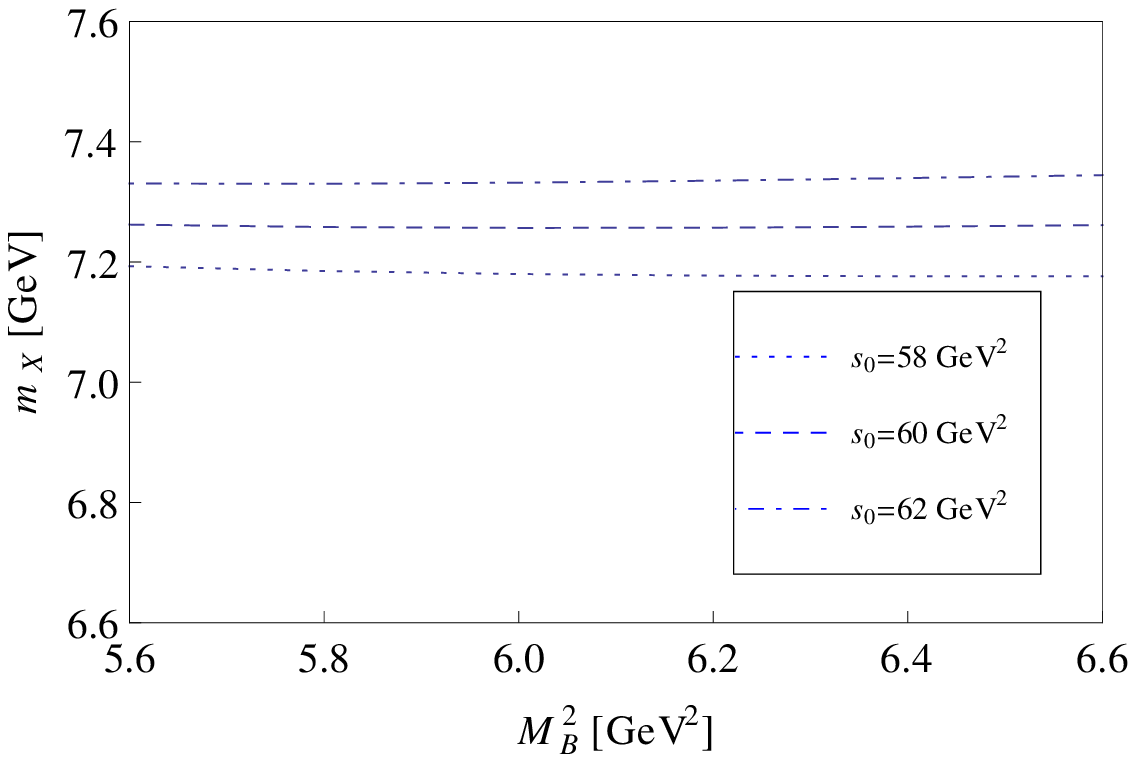}}
\end{tabular}
\figcaption{Variation of $m_X$ with $s_0$ and $M^2_B$ corresponding
to the current $J_4$ for the $0^{+}$ $bc\bar s\bar s$ system.}
\label{figbcss0+4}
\end{center}
\begin{center}
\renewcommand{\arraystretch}{1.3}
\begin{tabular*}{13cm}{cccccccc}
\hlinewd{.8pt}
System            & Current  & $s_0 (\mbox{GeV}^2)$ & $[M^2_{\mbox{min}}$,$M^2_{\mbox{max}}] (\mbox{GeV}^2)$ & $m_X$\mbox{(GeV)}&PC(\%)\\
\hline
$bc\bar q\bar q$   & $J_{1}$   &  $60\pm2$               & $5.4-6.2$                   & $7.27\pm0.08\pm0.06\pm0.05$    & 35.5  \\
                   & $J_{2}$   &  $59\pm2$               & $6.1-6.4$                   & $7.16\pm0.09\pm0.06\pm0.01$    & 32.9  \\
                   & $J_{3}$   &  $58\pm2$               & $5.4-6.0$                   & $7.14\pm0.08\pm0.05\pm0.03$    & 33.9  \\
                   & $J_{4}$   &  $60\pm2$               & $6.1-6.4$                   & $7.23\pm0.08\pm0.05\pm0.06$    & 33.5
\vspace{5pt}\\
$bc\bar s\bar s$   & $J_{1}$   &  $61\pm2$               & $4.9-6.4$                   & $7.35\pm0.08\pm0.06\pm0.03$    & 39.1  \\
                   & $J_{4}$   &  $60\pm2$               & $5.6-6.5$                   & $7.26\pm0.08\pm0.06\pm0.10$    & 36.7   \\
\hlinewd{1.0pt}
\end{tabular*}
\tabcaption{The threshold value, Borel window, mass and pole
contribution for the $J^P=0^+$ $bc\bar q\bar q$ and $bc\bar s\bar s$
systems.\label{tablebcqq0+}}
\end{center}


\begin{center}
\renewcommand{\arraystretch}{1.3}
\begin{tabular*}{13cm}{cccccccc}
\hlinewd{.8pt} System  & Current  & $s_0 (\mbox{GeV}^2)$ & $[M^2_{\mbox{min}}$,$M^2_{\mbox{max}}] (\mbox{GeV}^2)$ & $m_X$
\mbox{(GeV)}&PC(\%)\\
\hline
$bc\bar q\bar q$   & $J_{1\mu}$   &  $59\pm2$               & $5.5-6.1$                  & $7.21\pm0.08\pm0.05\pm0.03$    & 34.7  \\
                   & $J_{2\mu}$   &  $60\pm2$               & $5.3-6.2$                  & $7.27\pm0.09\pm0.06\pm0.05$    & 37.5  \\
                   & $J_{3\mu}$   &  $60\pm2$               & $5.4-6.3$                  & $7.26\pm0.08\pm0.06\pm0.05$    & 36.8  \\
                   & $J_{4\mu}$   &  $58\pm2$               & $5.3-6.0$                  & $7.13\pm0.08\pm0.06\pm0.03$    & 35.7
\vspace{5pt}\\
$bc\bar s\bar s$   & $J_{2\mu}$   &  $61\pm2$               & $4.9-6.4$                  & $7.35\pm0.07\pm0.11\pm0.04$    & 41.2  \\
                   & $J_{3\mu}$   &  $61\pm2$               & $4.9-6.4$                  & $7.34\pm0.07\pm0.07\pm0.08$    & 42.1   \\
\hlinewd{1.0pt}
\end{tabular*}
\tabcaption{The threshold value, Borel window, mass and pole
contribution for $J^P=1^+$ $bc\bar q\bar q$ and $bc\bar s\bar s$
systems.
\label{tablebcqq1+}}
\end{center}

The above analyses can easily be extended to the $bc\bar s\bar s$
systems by replacing the corresponding parameters such as the light
quark mass and various condensates. We expand the spectral densities
to first order in $m_s$ because $m_s$ is much larger than $m_q$ and
thus cannot be omitted.
These terms are very important to the OPE convergence and the mass sum
rule stability for the $bc\bar s\bar s$ systems. As mentioned in
Sec.~\ref{sec:current}, only $J_1$, $J_4$ with $J^P=0^+$ in
Eq.~\eqref{current1} and $J_{2\mu}$, $J_{3\mu}$ with $J^P=1^+$ in
Eq.~\eqref{current2} survive in the $bc\bar s\bar s$ system.
For the
currents $J_4$ with $J^P=0^+$, we show the variation of the
extracted mass $m_X$ with the threshold value $s_0$ and Borel
parameter $M^2_B$ in Fig.~\ref{figbcss0+4}. We obtain the threshold
value $s_0=60$ GeV$^2$ and the Borel window $5.6$ GeV$^2$ $\leq
M_B^2\leq 6.5$ GeV$^2$. Compared to the $bc\bar q\bar q$ system, the
Borel window of the $bc\bar s\bar s$ system becomes broader
because the pole contribution of the $bc\bar s\bar s$ channel is
larger than that of the $bc\bar q\bar q$ channel and the OPE
convergence becomes better. Finally, we extract the hadron mass
around $m_X=7.26\pm0.08\pm0.06\pm0.10$ GeV. After performing the
numerical analyses for all currents, we list the numerical results of
the $0^+$ $bc\bar s\bar s$ system in Table~\ref{tablebcqq0+} and the
$1^+$ $bc\bar s\bar s$ system in Table~\ref{tablebcqq1+}. For the
same current and QCD input parameters, the extracted mass of the
$bc\bar s\bar s$ state is about $0.1$ GeV higher than that of the
$bc\bar q\bar q$ state.

\subsection{$qc\bar q\bar b$ and $sc\bar s\bar b$ tetraquark systems}
In this subsection, we study $qc\bar q\bar b$ and $sc\bar s\bar b$
tetraquark systems with $J^P=0^+, 1^+$. These configurations are
very different from the $bc\bar q\bar q$ and $bc\bar s\bar s$
tetraquark systems. In the correlation functions and the spectral
densities, the quark condensate $\qq$ and the quark gluon mixed
condensate $\qGqa$ contain terms proportional to the heavy quark
masses and they cannot be ignored. They give the most important
nonperturbative contributions to the correlation functions.
In particular, the quark condensate $\qq$ term is now the
dominant power correction to the correlation function.

To ensure OPE convergence, we require that the perturbative term
be larger than 3 times of the quark condensate to obtain a
lower bound on the Borel parameter. Requiring PC be larger than 
$10\%$ leads to an upper bound on $M_B^2$. After studying the pole 
contribution, we find that the PC in all channels for the 
$qc\bar q\bar b$ and $sc\bar s\bar b$
tetraquark systems are much smaller than those for the $bc\bar q\bar
q$ and $bc\bar s\bar s$ tetraquark systems. This means that the
Borel windows in the $qc\bar q\bar b$ and $sc\bar s\bar b$ systems
will be much narrower than those in the $bc\bar q\bar q$ and $bc\bar
s\bar s$ systems.

\begin{center}
\begin{tabular}{lr}
\scalebox{0.65}{\includegraphics{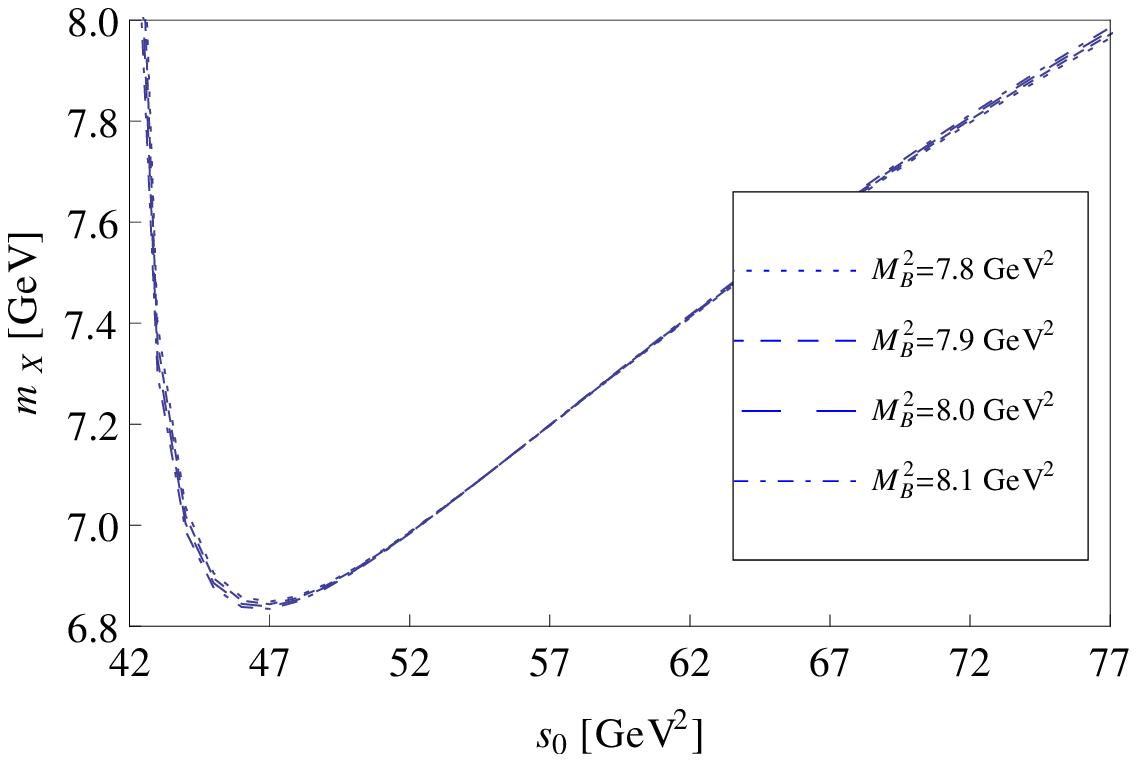}}&
\scalebox{0.65}{\includegraphics{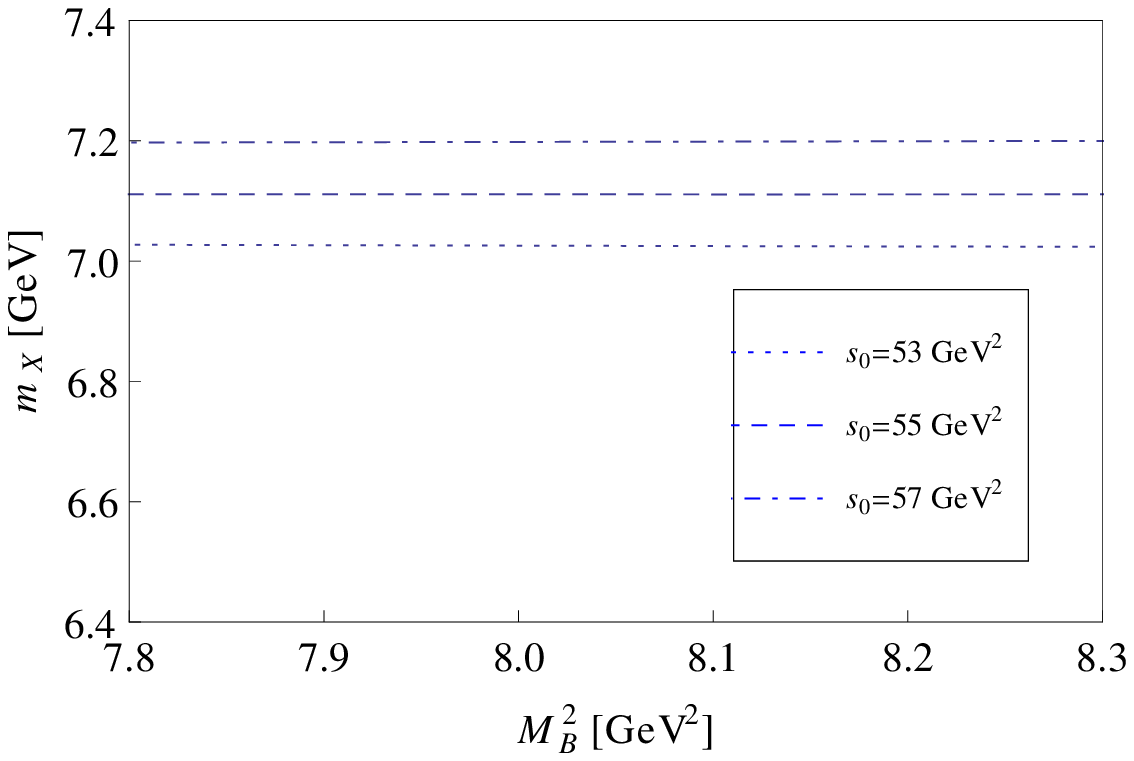}}
\end{tabular}
\figcaption{Variation of $m_X$ with $s_0$ and $M^2_B$ corresponding
to the current $J_1$ for the $0^{+}$ $qc\bar q\bar b$ system.}
\label{figcqbq0+3}
\end{center}

For the $J^P=0^+$ $qc\bar q\bar b$ system, only the current $J_1$
gives a significant (although narrow) Borel window under the above criteria.
The pole contributions of the currents $J_2, J_3$ and $J_4$ are too
small to give a suitable working region of the Borel mass. In
Fig.~\ref{figcqbq0+3}, we show the Borel curves of the extracted
mass with the threshold value $s_0$ and the Borel parameter $M_B^2$
using the interpolating current $J_1$. For $s_0=55$ GeV$^2$, we
obtain a very narrow Borel window $7.8$ GeV$^2$ $\leq M_B^2\leq 8.0$
GeV$^2$. In this region, the mass sum rule is very stable and the
hadron mass is finally extracted as $m_X=7.11$ GeV.

However, the $sc\bar s\bar b$ systems are much better. The interpolating
currents $J_1, J_2$ and $J_4$ can result in stable mass sum rules and allow
reliable extraction of hadron masses. In Fig.~\ref{figcsbs0+3}, we show the
Borel curves for the current
$J_1$ in the $sc\bar s\bar b$ system. For $s_0=56$ GeV$^2$, the
Borel window is determined as $6.6$ GeV$^2$ $\leq M_B^2\leq 8.1$
GeV$^2$, which is much broader than the corresponding $qc\bar q\bar
b$ system for the same current $J_1$. In the expressions \eqref{eq1}--\eqref{eq2},
the order $m_s$ parts in the perturbative and quark condensate terms
have opposite signs, enhancing the strange quark contributions and
resulting in a smaller lower bound on $M_B^2$.
This is the reason that the OPE convergence
of the $sc\bar s\bar b$ system is better than that of the $qc\bar q\bar b$
system.

\begin{center}
\begin{tabular}{lr}
\scalebox{0.65}{\includegraphics{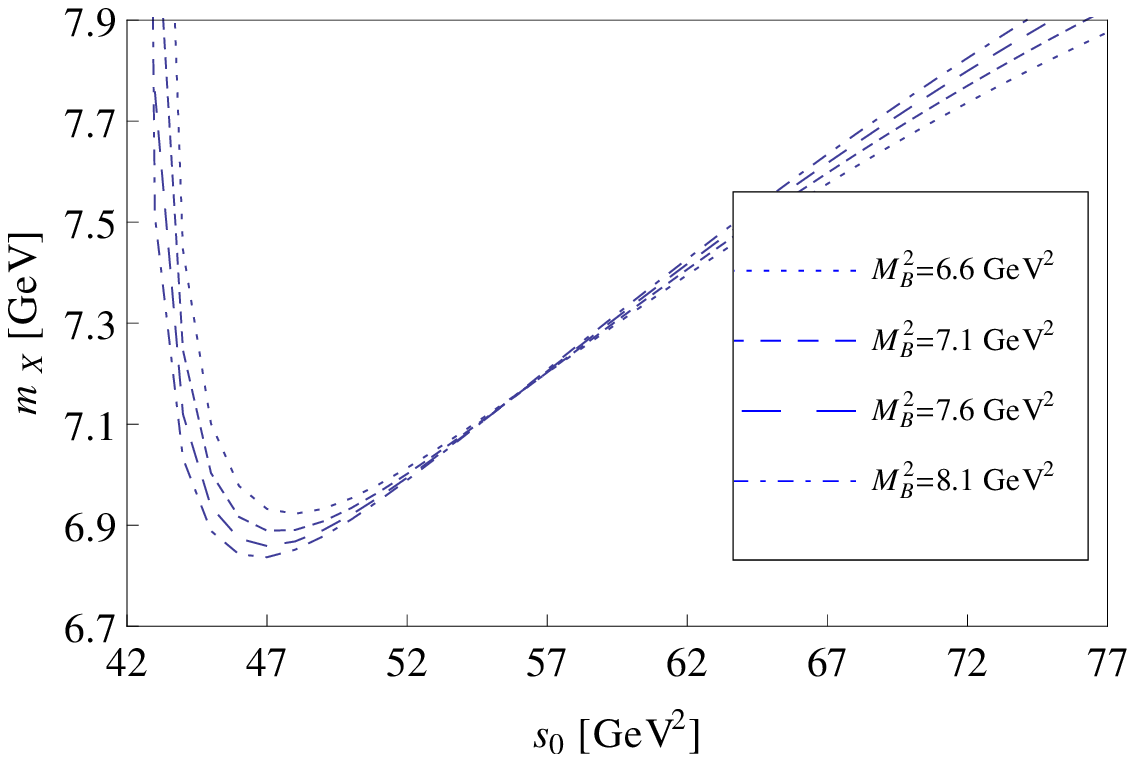}}&
\scalebox{0.65}{\includegraphics{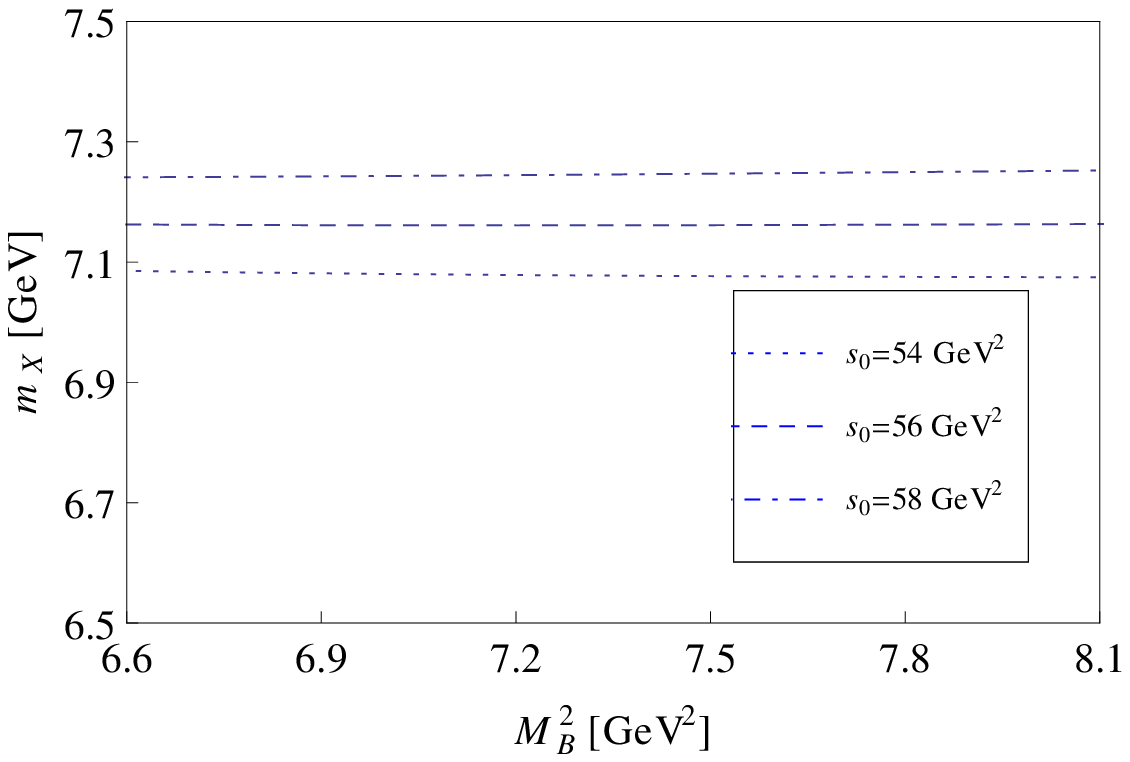}}
\end{tabular}
\figcaption{Variation of $m_X$ with $s_0$ and $M^2_B$ corresponding
to the current $J_1$ for the $0^{+}$ $sc\bar s\bar b$ system.}
\label{figcsbs0+3}
\end{center}
\begin{center}
\renewcommand{\arraystretch}{1.3}
\begin{tabular*}{13cm}{cccccccc}
\hlinewd{.8pt} System            & Current  & $s_0 (\mbox{GeV}^2)$
& $[M^2_{\mbox{min}}$,$M^2_{\mbox{max}}] (\mbox{GeV}^2)$ &
$m_X$\mbox{(GeV)}&PC(\%)\\
\hline
$qc\bar q\bar b$   & $J_{1}$   &  $55\pm2$               &$7.8-8.0$                     &$7.11\pm0.08\pm0.06\pm0.01$    & 10.2
\vspace{5pt}\\
$sc\bar s\bar b$   & $J_{1}$   &  $56\pm2$               & $6.6-8.1$                    &$7.16\pm0.08\pm0.06\pm0.04$    & 14.4  \\
                   & $J_{2}$   &  $56\pm2$               & $8.8-9.2$                    &$7.10\pm0.09\pm0.04\pm0.13$    & 10.6  \\
                   & $J_{4}$   &  $56\pm2$               & $8.8-9.1$                    &$7.10\pm0.09\pm0.06\pm0.12$    & 10.9   \\
\hlinewd{1.0pt}
\end{tabular*}
\tabcaption{The threshold value, Borel window, mass and pole
contribution for $J^P=0^+$ $qc\bar q\bar b$ and $sc\bar s\bar b$
systems.
\label{tableqcqb0+}}
\end{center}

\begin{center}
\renewcommand{\arraystretch}{1.3}
\begin{tabular*}{13cm}{cccccccc}
\hlinewd{.8pt}
System             & Current  & $s_0 (\mbox{GeV}^2)$ & $[M^2_{\mbox{min}}$,$M^2_{\mbox{max}}] (\mbox{GeV}^2)$ & $m_X$\mbox{(GeV)}&PC(\%)\\
\hline
$qc\bar q\bar b$   & $J_{1\mu}$   &  $55\pm2$               & $7.9-8.2$                   & $7.10\pm0.09\pm0.06\pm0.01$    & 10.4  \\
                   & $J_{2\mu}$   &  $55\pm2$               & $7.9-8.2$                   & $7.09\pm0.09\pm0.06\pm0.01$    & 10.7
\vspace{5pt}\\
$sc\bar s\bar b$   & $J_{1\mu}$   &  $55\pm2$               & $6.7-7.9$                   & $7.11\pm0.08\pm0.05\pm0.03$    & 14.0  \\
                   & $J_{2\mu}$   &  $56\pm2$               & $6.7-8.3$                   & $7.15\pm0.09\pm0.06\pm0.05$    & 14.2  \\
                   & $J_{3\mu}$   &  $52\pm2$               & $6.7-7.3$                   & $6.90\pm0.09\pm0.02\pm0.03$    & 11.6  \\
                   & $J_{4\mu}$   &  $52\pm2$               & $6.7-7.3$                   & $6.92\pm0.09\pm0.06\pm0.03$    & 11.0  \\
\hlinewd{1.0pt}
\end{tabular*}
\tabcaption{The threshold value, Borel window, mass and pole
contribution for $J^P=1^+$ $qc\bar q\bar b$ and $sc\bar s\bar b$
systems.
\label{tableqcqb1+}}
\end{center}

We collect the numerical results for the scalar and axial vector
$qc\bar q\bar b$ systems in Tables~\ref{tableqcqb0+} and
\ref{tableqcqb1+} respectively, including the continuum threshold
values, the Borel windows, the extracted masses and the pole
contributions.

As mentioned above, the pole contributions of these $qc\bar q\bar b$
and $sc\bar s\bar b$ systems are very small making it difficult to
obtain a significant Borel window. To improve the pole contribution
and sum rule reliability, one possible way is using the
mixed interpolating currents to calculate the spectral densities and
correlation functions~\cite{2007-Chen-p94025-94025}.
For both the $J^P=0^+$ and $1^+$ $qc\bar
q\bar b$ systems, $J_1$ and $J_3$ have similar Lorentz structures,
which result in very similar spectral densities in the 
Appendix. The same situation exists for $J_2$ and
$J_4$. So the reasonable choice is mixing $J_1$ with $J_2$ and mixing
$J_3$ with $J_4$. However, these two mixed currents will also give
the similar results due to their Lorentz structures. We therefore
consider the following mixed currents:
\begin{eqnarray}
J^m=\cos\theta J_1+\sin\theta J_2, \label{mixcurrent1}
\end{eqnarray}
for $J^P=0^+$ $qc\bar q\bar b$ system and
\begin{eqnarray}
J^m_{\mu}=\cos\theta J_{1\mu}+\sin\theta J_{2\mu},
\label{mixcurrent2}
\end{eqnarray}
for $J^P=1^+$ $qc\bar q\bar b$ system.

For $J^m$ and $J_{\mu}^m$, we just need to calculate the mixed parts
$\langle 0|T[J_1J_2^{\dag}]|0\rangle+\langle
0|T[J_2J_1^{\dag}]|0\rangle$ and $\langle
0|T[J_{1\mu}J_{2\nu}^{\dag}]|0\rangle+\langle
0|T[J_{2\mu}J_{1\nu}^{\dag}]|0\rangle$ in the correlation functions.
In the Appendix, we list the spectral densities of these
two mixed parts. In these expressions, the perturbative terms, the
quark condensate and the four quark condensate give no contributions
to the correlation functions. Utilizing these results and the
spectral densities for $J_{1\mu}$ and $J_{2\mu}$, we perform the
numerical analysis in the axial vector $qc\bar q\bar b$ channel with
the mixed current $J_{\mu}^m$. Under the same criteria of the OPE
convergence and pole contribution, we obtain the Borel window $7.9$
GeV$^2$ $\leq M_B^2\leq 8.4$ GeV$^2$ with $s_0=55$ GeV$^2$. To study
the mixing effect, we show the variation of the pole contribution
with the mixing angle $\theta$ in Fig.~\ref{figmixPC}. It shows that
there is no significant enhancement of the pole contribution for all
the value of mixing angle. In Fig.~\ref{figcqbqmix1+}, we show the
Borel curves of the extracted mass with $s_0$ and $M_B^2$ for the 
$J^P=1^{+}$ $qc\bar q\bar b$ system with the mixed current
$J_{\mu}^m$. Finally, we extract the ground state mass around $7.11$
GeV. Compared to the numerical results from the single current in
Table~\ref{tableqcqb1+}, the mass, continuum threshold, Borel window
and pole contribution from the mixed current $J_{\mu}^m$ are almost
the same. The similar situation occurs for the mixed current  $J^m$.
In other words, the mixed current does not improve the mass
sum rules significantly.
\begin{center}
\includegraphics[scale=0.75]{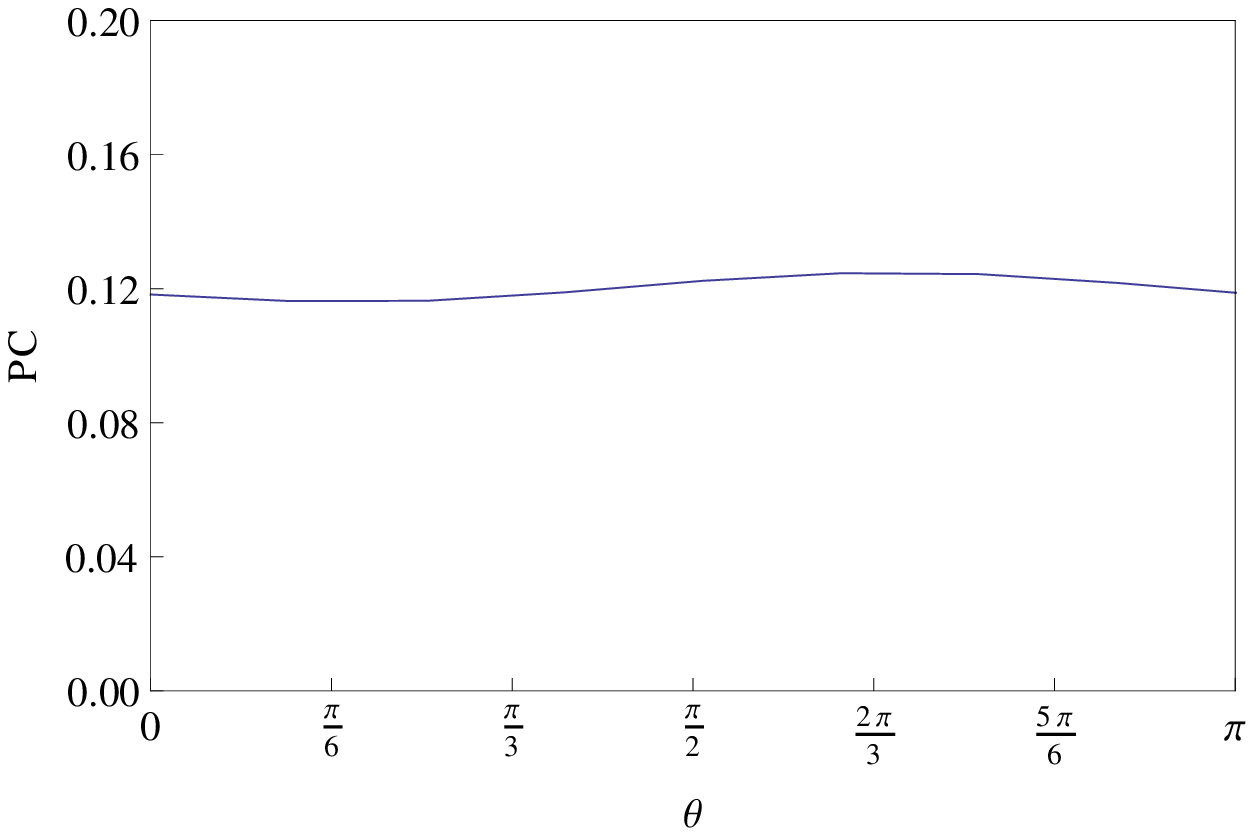}
\figcaption{Pole contribution as a function of the mixing angle
$\theta$ with $s_0=55$ GeV$^2$ and $M_B^2=8.0$ GeV$^2$ for
$J_{\mu}^m$.} \label{figmixPC}
\end{center}
\begin{center}
\begin{tabular}{lr}
\scalebox{0.65}{\includegraphics{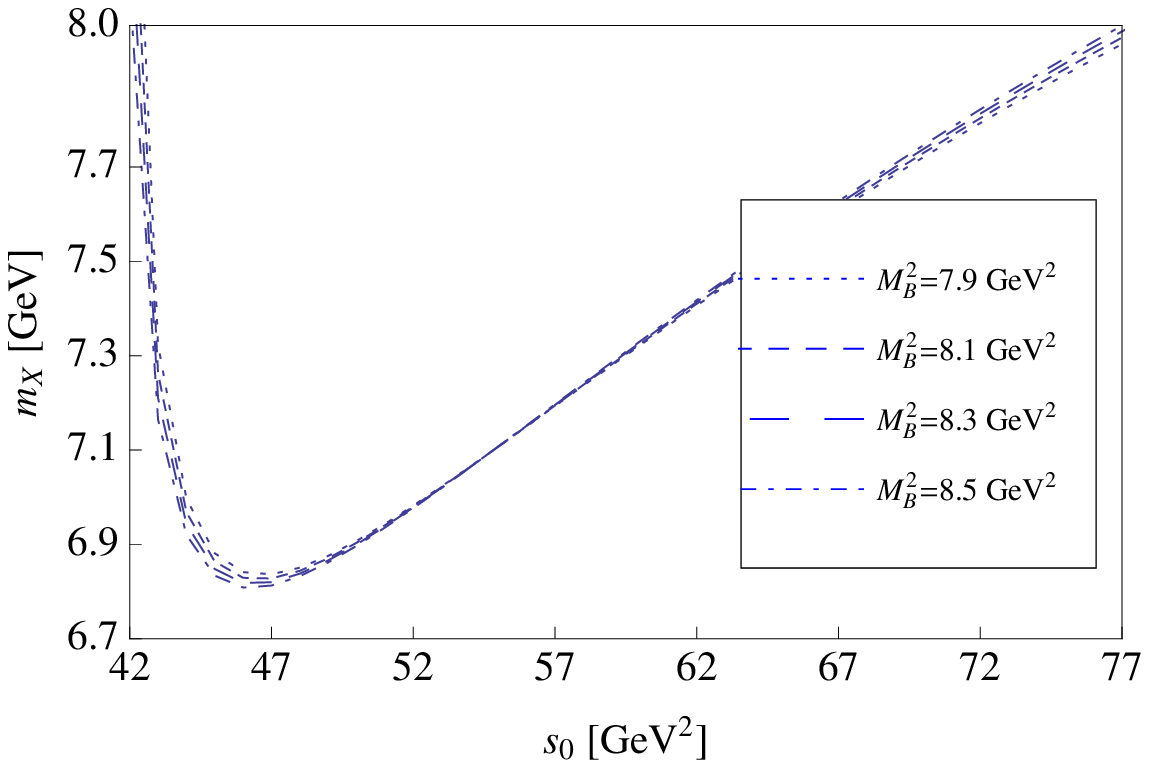}}&
\scalebox{0.65}{\includegraphics{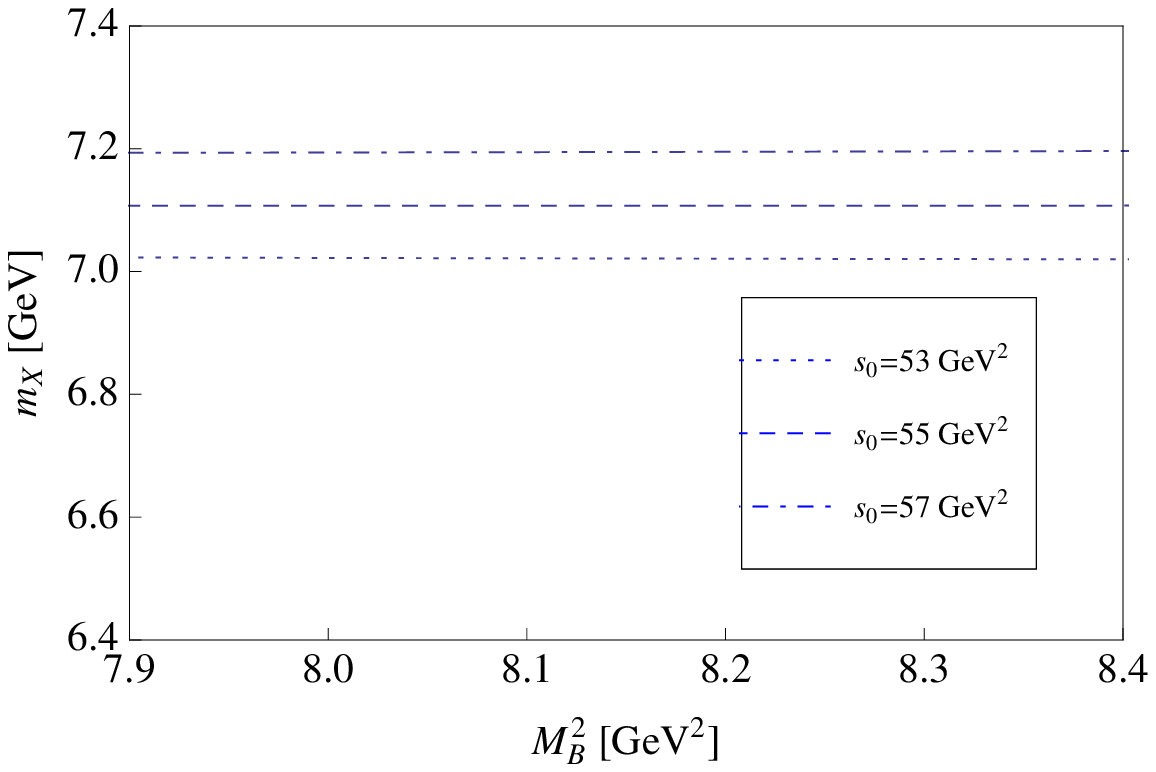}}
\end{tabular}
\figcaption{Variation of $m_X$ with $s_0$ and $M^2_B$ corresponding
to the mixed current $J_{\mu}^m$ for the $1^{+}$ $qc\bar q\bar b$
system.} \label{figcqbqmix1+}
\end{center}

\section{SUMMARY}\label{sec:SUMMARY}
We have constructed the $bc\bar{q}\bar{q}$, $bc\bar{s}\bar{s}$ and
$qc\bar{q}\bar{b}$, $sc\bar{s}\bar{b}$ tetraquark currents with
$J^P=0^+$ and $1^+$. At the leading order in $\alpha_s$, we
calculate the two-point correlation functions and the spectral
densities including the contributions of the perturbative terms,
quark condensate $\qq$, gluon condensate $\GGa$, quark-gluon mixed
condensate $\qGqa$, four quark condensate $\qq^2$ and dimension
eight condensate $\qq\qGqa$.

For the $bc\bar{q}\bar{q}$ systems, both the quark condensate $\qq$
and quark-gluon mixed condensate $\qGqa$ are proportional to the
light quark mass $m_q$ and vanish in the chiral limit $m_q=0$. The
four quark condensate $\qq^2$ is the dominant power correction to
the correlation functions. The dimension eight condensate $\qq\qGqa$
also gives an important contribution. To study the $bc\bar{s}\bar{s}$
systems, we keep the leading-order $m_s$ corrections to the spectral
densities. The numerical analysis shows that these terms can improve
the OPE convergence and pole contribution to enlarge the Borel window
of the mass sum rules. The extracted masses for both the scalar and
axial vector $bc\bar{q}\bar{q}$ and $bc\bar{s}\bar{s}$ tetraquark states
are about $7.1-7.2$ GeV and $7.2-7.3$ GeV, respectively.

The situation for the $qc\bar{q}\bar{b}$ systems is very different from
that of the $bc\bar{q}\bar{q}$ systems. The quark condensate $\qq$ and
quark-gluon mixed condensate $\qGqa$ are multiplied by the heavy quark
mass $m_Q$ and give important contributions to the correlation functions.
The quark condensate is the dominant
power correction in these systems. After performing the numerical
analysis, we extract the masses of both the scalar and axial vector
$qc\bar{q}\bar{b}$ states around $7.1$ GeV. The mass is about $7.1$
GeV for the scalar $sc\bar{s}\bar{b}$ state and $6.9-7.1$ GeV for
the axial vector $sc\bar{s}\bar{b}$ state. However, the pole
contributions of these $qc\bar{q}\bar{b}$ systems are so small that
the corresponding Borel windows are very narrow. To improve the pole
contributions and enlarge the Borel windows, we investigated the
mixed interpolating currents by introducing a mixing angle $\theta$.
Unfortunately, the numerical analysis shows that these mixed currents
give no significant effects that would expand the Borel window.

The masses of these $bc\bar{q}\bar{q}$, $bc\bar{s}\bar{s}$ and $qc\bar{q}\bar{b}$,
$sc\bar{s}\bar{b}$ tetraquark states are below the open-flavor thresholds $D^{(\ast)}\bar B^{(\ast)}$,
$D_s^{(\ast)}\bar B_s^{(\ast)}$ and $D^{(\ast)}B^{(\ast)}$, $D_s^{(\ast)}B_s^{(\ast)}$, respectively.
In other words, these tetraquark states $bc\bar{q}\bar{q}$, $bc\bar{s}\bar{s}$
and $qc\bar{q}\bar{b}$, $sc\bar{s}\bar{b}$ cannot decay into the
open-flavor modes due to the kinematics limits. On the other hand,
the $B_c$ plus light meson decay modes for the $qc\bar{q}\bar{b}$
states are allowed, such as $X (0^+)\to B_c\pi, B_c\eta$ and
$X (1^+)\to B_c\rho, B_c\omega$. Such channels are suggested for
the future search of these possible $qc\bar{q}\bar{b}$, $sc\bar{s}\bar{b}$
states. The $bc\bar{q}\bar{q}$ and $bc\bar{s}\bar{s}$ tetraquark states
cannot decay through these fall-apart mechanisms, suggesting dominantly
weak decay mechanisms. They may be produced at facilities such as
Super-B factories, LHCb, PANDA and RHIC.

\section*{ACKNOWLEDGMENTS}

This project was supported by the Natural Sciences and Engineering
Research Council of Canada (NSERC). S. L. Z. was supported by the
National Natural Science Foundation of China under Grant NO. 
11261130311.



\begin{thebibliography}{55}
\expandafter\ifx\csname
natexlab\endcsname\relax\def\natexlab#1{#1}\fi
\expandafter\ifx\csname bibnamefont\endcsname\relax
  \def\bibnamefont#1{#1}\fi
\expandafter\ifx\csname bibfnamefont\endcsname\relax
  \def\bibfnamefont#1{#1}\fi
\expandafter\ifx\csname citenamefont\endcsname\relax
  \def\citenamefont#1{#1}\fi
\expandafter\ifx\csname url\endcsname\relax
  \def\url#1{\texttt{#1}}\fi
\expandafter\ifx\csname urlprefix\endcsname\relax\def\urlprefix{URL
}\fi \providecommand{\bibinfo}[2]{#2}
\providecommand{\eprint}[2][]{\url{#2}}

\bibitem[{\citenamefont{Klempt and Zaitsev}(2007)}]{2007-Klempt-p1-202}
\bibinfo{author}{\bibfnamefont{E.}~\bibnamefont{Klempt}} \bibnamefont{and}
  \bibinfo{author}{\bibfnamefont{A.}~\bibnamefont{Zaitsev}},
  \bibinfo{journal}{Phys. Rep.} \textbf{\bibinfo{volume}{454}},
  \bibinfo{pages}{1} (\bibinfo{year}{2007}).

\bibitem[{\citenamefont{Beringer et~al.}(2012)}]{2012-Beringer-p10001-10001}
\bibinfo{author}{\bibfnamefont{J.}~\bibnamefont{Beringer}} \bibnamefont{et~al.}
  (\bibinfo{collaboration}{Particle Data Group}), \bibinfo{journal}{Phys.Rev.}
  \textbf{\bibinfo{volume}{D86}}, \bibinfo{pages}{010001}
  (\bibinfo{year}{2012}).

\bibitem[{\citenamefont{Jaffe}(1977{\natexlab{a}})}]{1977-Jaffe-p267-267}
\bibinfo{author}{\bibfnamefont{R.~L.} \bibnamefont{Jaffe}},
  \bibinfo{journal}{Phys. Rev.} \textbf{\bibinfo{volume}{D15}},
  \bibinfo{pages}{267} (\bibinfo{year}{1977}{\natexlab{a}}).

\bibitem[{\citenamefont{Jaffe}(1977{\natexlab{b}})}]{1977-Jaffe-p281-281}
\bibinfo{author}{\bibfnamefont{R.~L.} \bibnamefont{Jaffe}},
  \bibinfo{journal}{Phys. Rev.} \textbf{\bibinfo{volume}{D15}},
  \bibinfo{pages}{281} (\bibinfo{year}{1977}{\natexlab{b}}).

\bibitem[{\citenamefont{Chen et~al.}(2007{\natexlab{a}})\citenamefont{Chen,
  Hosaka, and Zhu}}]{2007-Chen-p94025-94025}
\bibinfo{author}{\bibfnamefont{H.-X.} \bibnamefont{Chen}},
  \bibinfo{author}{\bibfnamefont{A.}~\bibnamefont{Hosaka}}, \bibnamefont{and}
  \bibinfo{author}{\bibfnamefont{S.-L.} \bibnamefont{Zhu}},
  \bibinfo{journal}{Phys. Rev.} \textbf{\bibinfo{volume}{D76}},
  \bibinfo{pages}{094025} (\bibinfo{year}{2007}{\natexlab{a}}).

\bibitem[{\citenamefont{Chen et~al.}(2007{\natexlab{b}})\citenamefont{Chen,
  Hosaka, and Zhu}}]{2007-Chen-p369-372}
\bibinfo{author}{\bibfnamefont{H.-X.} \bibnamefont{Chen}},
  \bibinfo{author}{\bibfnamefont{A.}~\bibnamefont{Hosaka}}, \bibnamefont{and}
  \bibinfo{author}{\bibfnamefont{S.-L.} \bibnamefont{Zhu}},
  \bibinfo{journal}{Phys. Lett.} \textbf{\bibinfo{volume}{B650}},
  \bibinfo{pages}{369} (\bibinfo{year}{2007}{\natexlab{b}}),
  \eprint{hep-ph/0609163}.


\bibitem[{\citenamefont{Zhang et~al.}(2007)\citenamefont{Zhang, Huang, and
  Steele}}]{2007-Zhang-p36004-36004}
\bibinfo{author}{\bibfnamefont{A.}~\bibnamefont{Zhang}},
  \bibinfo{author}{\bibfnamefont{T.}~\bibnamefont{Huang}}, \bibnamefont{and}
  \bibinfo{author}{\bibfnamefont{T.~G.}~\bibnamefont{Steele}},
  \bibinfo{journal}{Phys.Rev.} \textbf{\bibinfo{volume}{D76}},
  \bibinfo{pages}{036004} (\bibinfo{year}{2007}), \eprint{hep-ph/0612146}.

\bibitem[{\citenamefont{Ebert et~al.}(2006)\citenamefont{Ebert, Faustov, and
  Galkin}}]{2006-Ebert-p214-219}
\bibinfo{author}{\bibfnamefont{D.}~\bibnamefont{Ebert}},
  \bibinfo{author}{\bibfnamefont{R.~N.} \bibnamefont{Faustov}},
  \bibnamefont{and} \bibinfo{author}{\bibfnamefont{V.~O.}
  \bibnamefont{Galkin}}, \bibinfo{journal}{Phys. Lett.}
  \textbf{\bibinfo{volume}{B634}}, \bibinfo{pages}{214} (\bibinfo{year}{2006}),
  \eprint{hep-ph/0512230}.

\bibitem[{\citenamefont{Ebert et~al.}(2008)\citenamefont{Ebert, Faustov, and
  Galkin}}]{2008-Ebert-p399-405}
\bibinfo{author}{\bibfnamefont{D.}~\bibnamefont{Ebert}},
  \bibinfo{author}{\bibfnamefont{R.~N.} \bibnamefont{Faustov}},
  \bibnamefont{and} \bibinfo{author}{\bibfnamefont{V.~O.}
  \bibnamefont{Galkin}}, \bibinfo{journal}{Eur. Phys. J.}
  \textbf{\bibinfo{volume}{C58}}, \bibinfo{pages}{399} (\bibinfo{year}{2008}).

\bibitem[{\citenamefont{Matheus et~al.}(2007)\citenamefont{Matheus, Narison,
  Nielsen, and Richard}}]{2007-Matheus-p14005-14005}
\bibinfo{author}{\bibfnamefont{R.~D.} \bibnamefont{Matheus}},
  \bibinfo{author}{\bibfnamefont{S.}~\bibnamefont{Narison}},
  \bibinfo{author}{\bibfnamefont{M.}~\bibnamefont{Nielsen}}, \bibnamefont{and}
  \bibinfo{author}{\bibfnamefont{J.~M.} \bibnamefont{Richard}},
  \bibinfo{journal}{Phys. Rev.} \textbf{\bibinfo{volume}{D75}},
  \bibinfo{pages}{014005} (\bibinfo{year}{2007}), \eprint{hep-ph/0608297}.

\bibitem[{\citenamefont{Bracco et~al.}(2009)\citenamefont{Bracco, Lee, Nielsen,
  and Rodrigues~da Silva}}]{2009-Bracco-p240-244}
\bibinfo{author}{\bibfnamefont{M.~E.} \bibnamefont{Bracco}},
  \bibinfo{author}{\bibfnamefont{S.~H.} \bibnamefont{Lee}},
  \bibinfo{author}{\bibfnamefont{M.}~\bibnamefont{Nielsen}}, \bibnamefont{and}
  \bibinfo{author}{\bibfnamefont{R.}~\bibnamefont{Rodrigues~da Silva}},
  \bibinfo{journal}{Phys. Lett.} \textbf{\bibinfo{volume}{B671}},
  \bibinfo{pages}{240} (\bibinfo{year}{2009}).

\bibitem[{\citenamefont{Chen and Zhu}(2010)}]{2010-Chen-p105018-105018}
\bibinfo{author}{\bibfnamefont{W.}~\bibnamefont{Chen}} \bibnamefont{and}
  \bibinfo{author}{\bibfnamefont{S.-L.} \bibnamefont{Zhu}},
  \bibinfo{journal}{Phys. Rev.} \textbf{\bibinfo{volume}{D81}},
  \bibinfo{pages}{105018} (\bibinfo{year}{2010}).

\bibitem[{\citenamefont{Chen and Zhu}(2011)}]{2011-Chen-p34010-34010}
\bibinfo{author}{\bibfnamefont{W.}~\bibnamefont{Chen}} \bibnamefont{and}
  \bibinfo{author}{\bibfnamefont{S.-L.} \bibnamefont{Zhu}},
  \bibinfo{journal}{Phys. Rev.} \textbf{\bibinfo{volume}{D83}},
  \bibinfo{pages}{034010} (\bibinfo{year}{2011}).

\bibitem[{\citenamefont{Du et~al.}(2013{\natexlab{a}})\citenamefont{Du, Chen,
  Chen, and Zhu}}]{2013-Du-p33104-33104}
\bibinfo{author}{\bibfnamefont{M.-L.} \bibnamefont{Du}},
  \bibinfo{author}{\bibfnamefont{W.}~\bibnamefont{Chen}},
  \bibinfo{author}{\bibfnamefont{X.-L.} \bibnamefont{Chen}}, \bibnamefont{and}
  \bibinfo{author}{\bibfnamefont{S.-L.} \bibnamefont{Zhu}},
  \bibinfo{journal}{Chin.Phys.} \textbf{\bibinfo{volume}{C37}},
  \bibinfo{pages}{033104} (\bibinfo{year}{2013}{\natexlab{a}}).

\bibitem[{\citenamefont{Maiani et~al.}(2005)\citenamefont{Maiani, Riquer,
  Piccinini, and Polosa}}]{2005-Maiani-p31502-31502}
\bibinfo{author}{\bibfnamefont{L.}~\bibnamefont{Maiani}},
  \bibinfo{author}{\bibfnamefont{V.}~\bibnamefont{Riquer}},
  \bibinfo{author}{\bibfnamefont{F.}~\bibnamefont{Piccinini}},
  \bibnamefont{and} \bibinfo{author}{\bibfnamefont{A.~D.}
  \bibnamefont{Polosa}}, \bibinfo{journal}{Phys. Rev.}
  \textbf{\bibinfo{volume}{D72}}, \bibinfo{pages}{031502}
  (\bibinfo{year}{2005}), \eprint{hep-ph/0507062}.

\bibitem[{\citenamefont{Maiani et~al.}(2007)\citenamefont{Maiani, Polosa, and
  Riquer}}]{2007-Maiani-p182003-182003}
\bibinfo{author}{\bibfnamefont{L.}~\bibnamefont{Maiani}},
  \bibinfo{author}{\bibfnamefont{A.~D.} \bibnamefont{Polosa}},
  \bibnamefont{and} \bibinfo{author}{\bibfnamefont{V.}~\bibnamefont{Riquer}},
  \bibinfo{journal}{Phys. Rev. Lett.} \textbf{\bibinfo{volume}{99}},
  \bibinfo{pages}{182003} (\bibinfo{year}{2007}).

\bibitem[{\citenamefont{Maiani et~al.}(2008)\citenamefont{Maiani, Polosa, and
  Riquer}}]{2008-Maiani-p73004-73004}
\bibinfo{author}{\bibfnamefont{L.}~\bibnamefont{Maiani}},
  \bibinfo{author}{\bibfnamefont{A.~D.} \bibnamefont{Polosa}},
  \bibnamefont{and} \bibinfo{author}{\bibfnamefont{V.}~\bibnamefont{Riquer}},
  \bibinfo{journal}{New J. Phys.} \textbf{\bibinfo{volume}{10}},
  \bibinfo{pages}{073004} (\bibinfo{year}{2008}).

\bibitem[{\citenamefont{Kleiv et~al.}(2013)\citenamefont{Kleiv, Steele, Zhang,
  and Blokland}}]{2013-Kleiv-p125018-125018}
\bibinfo{author}{\bibfnamefont{R.~T.}~\bibnamefont{Kleiv}},
  \bibinfo{author}{\bibfnamefont{T.~G.}~\bibnamefont{Steele}},
  \bibinfo{author}{\bibfnamefont{A.}~\bibnamefont{Zhang}}, \bibnamefont{and}
  \bibinfo{author}{\bibfnamefont{I.}~\bibnamefont{Blokland}},
  \bibinfo{journal}{Phys.Rev.} \textbf{\bibinfo{volume}{D87}},
  \bibinfo{pages}{125018} (\bibinfo{year}{2013}).

\bibitem[{\citenamefont{Carlson et~al.}(1988)\citenamefont{Carlson, Heller, and
  Tjon}}]{1988-Carlson-p744-744}
\bibinfo{author}{\bibfnamefont{J.}~\bibnamefont{Carlson}},
  \bibinfo{author}{\bibfnamefont{L.}~\bibnamefont{Heller}}, \bibnamefont{and}
  \bibinfo{author}{\bibfnamefont{J.~A.}~\bibnamefont{Tjon}},
  \bibinfo{journal}{Phys.Rev.} \textbf{\bibinfo{volume}{D37}},
  \bibinfo{pages}{744} (\bibinfo{year}{1988}).

\bibitem[{\citenamefont{Zhang et~al.}(2008)\citenamefont{Zhang, Zhang, and
  Zhang}}]{2008-Zhang-p437-440}
\bibinfo{author}{\bibfnamefont{M.}~\bibnamefont{Zhang}},
  \bibinfo{author}{\bibfnamefont{H.}~\bibnamefont{Zhang}}, \bibnamefont{and}
  \bibinfo{author}{\bibfnamefont{Z.}~\bibnamefont{Zhang}},
  \bibinfo{journal}{Commun.Theor.Phys.} \textbf{\bibinfo{volume}{50}},
  \bibinfo{pages}{437} (\bibinfo{year}{2008}).

\bibitem[{\citenamefont{Pepin et~al.}(1997)\citenamefont{Pepin, Stancu,
  Genovese, and Richard}}]{1997-Pepin-p119-123}
\bibinfo{author}{\bibfnamefont{S.}~\bibnamefont{Pepin}},
  \bibinfo{author}{\bibfnamefont{F.}~\bibnamefont{Stancu}},
  \bibinfo{author}{\bibfnamefont{M.}~\bibnamefont{Genovese}}, \bibnamefont{and}
  \bibinfo{author}{\bibfnamefont{J.}~\bibnamefont{Richard}},
  \bibinfo{journal}{Phys.Lett.} \textbf{\bibinfo{volume}{B393}},
  \bibinfo{pages}{119} (\bibinfo{year}{1997}), \eprint{hep-ph/9609348}.

\bibitem[{\citenamefont{Vijande et~al.}(2006)\citenamefont{Vijande, Valcarce,
  and Tsushima}}]{2006-Vijande-p54018-54018}
\bibinfo{author}{\bibfnamefont{J.}~\bibnamefont{Vijande}},
  \bibinfo{author}{\bibfnamefont{A.}~\bibnamefont{Valcarce}}, \bibnamefont{and}
  \bibinfo{author}{\bibfnamefont{K.}~\bibnamefont{Tsushima}},
  \bibinfo{journal}{Phys.Rev.} \textbf{\bibinfo{volume}{D74}},
  \bibinfo{pages}{054018} (\bibinfo{year}{2006}), \eprint{hep-ph/0608316}.

\bibitem[{\citenamefont{Vijande et~al.}(2009)\citenamefont{Vijande, Valcarce,
  and Barnea}}]{2009-Vijande-p74010-74010}
\bibinfo{author}{\bibfnamefont{J.}~\bibnamefont{Vijande}},
  \bibinfo{author}{\bibfnamefont{A.}~\bibnamefont{Valcarce}}, \bibnamefont{and}
  \bibinfo{author}{\bibfnamefont{N.}~\bibnamefont{Barnea}},
  \bibinfo{journal}{Phys.Rev.} \textbf{\bibinfo{volume}{D79}},
  \bibinfo{pages}{074010} (\bibinfo{year}{2009}).

\bibitem[{\citenamefont{Brink and Stancu}(1998)}]{1998-Brink-p6778-6787}
\bibinfo{author}{\bibfnamefont{D.~M.}~\bibnamefont{Brink}} \bibnamefont{and}
  \bibinfo{author}{\bibfnamefont{Fl.}~\bibnamefont{Stancu}},
  \bibinfo{journal}{Phys.Rev.} \textbf{\bibinfo{volume}{D57}},
  \bibinfo{pages}{6778} (\bibinfo{year}{1998}).

\bibitem[{\citenamefont{Silvestre-Brac and
  Semay}(1993)}]{1993-Silvestre-Brac-p457-470}
\bibinfo{author}{\bibfnamefont{B.}~\bibnamefont{Silvestre-Brac}}
  \bibnamefont{and} \bibinfo{author}{\bibfnamefont{C.}~\bibnamefont{Semay}},
  \bibinfo{journal}{Z.Phys.} \textbf{\bibinfo{volume}{C59}},
  \bibinfo{pages}{457} (\bibinfo{year}{1993}).

\bibitem[{\citenamefont{Zouzou et~al.}(1986)\citenamefont{Zouzou,
  Silvestre-Brac, Gignoux, and Richard}}]{1986-Zouzou-p457-457}
\bibinfo{author}{\bibfnamefont{S.}~\bibnamefont{Zouzou}},
  \bibinfo{author}{\bibfnamefont{B.}~\bibnamefont{Silvestre-Brac}},
  \bibinfo{author}{\bibfnamefont{C.}~\bibnamefont{Gignoux}}, \bibnamefont{and}
  \bibinfo{author}{\bibfnamefont{J.}~\bibnamefont{Richard}},
  \bibinfo{journal}{Z.Phys.} \textbf{\bibinfo{volume}{C30}},
  \bibinfo{pages}{457} (\bibinfo{year}{1986}).

\bibitem[{\citenamefont{Ebert et~al.}(2007)\citenamefont{Ebert, Faustov,
  Galkin, and Lucha}}]{2007-Ebert-p114015-114015}
\bibinfo{author}{\bibfnamefont{D.}~\bibnamefont{Ebert}},
  \bibinfo{author}{\bibfnamefont{R.~N.}~\bibnamefont{Faustov}},
  \bibinfo{author}{\bibfnamefont{V.~O.}~\bibnamefont{Galkin}}, \bibnamefont{and}
  \bibinfo{author}{\bibfnamefont{W.}~\bibnamefont{Lucha}},
  \bibinfo{journal}{Phys.Rev.} \textbf{\bibinfo{volume}{D76}},
  \bibinfo{pages}{114015} (\bibinfo{year}{2007}).

\bibitem[{\citenamefont{Manohar and Wise}(1993)}]{1993-Manohar-p17-33}
\bibinfo{author}{\bibfnamefont{A.~V.} \bibnamefont{Manohar}} \bibnamefont{and}
  \bibinfo{author}{\bibfnamefont{M.~B.} \bibnamefont{Wise}},
  \bibinfo{journal}{Nucl.Phys.} \textbf{\bibinfo{volume}{B399}},
  \bibinfo{pages}{17} (\bibinfo{year}{1993}), \eprint{hep-ph/9212236}.

\bibitem[{\citenamefont{Cui et~al.}(2007)\citenamefont{Cui, Chen, Deng, and
  Zhu}}]{2007-Cui-p7-13}
\bibinfo{author}{\bibfnamefont{Y.}~\bibnamefont{Cui}},
  \bibinfo{author}{\bibfnamefont{X.-L.} \bibnamefont{Chen}},
  \bibinfo{author}{\bibfnamefont{W.-Z.} \bibnamefont{Deng}}, \bibnamefont{and}
  \bibinfo{author}{\bibfnamefont{S.-L.} \bibnamefont{Zhu}},
  \bibinfo{journal}{High Energy Phys. Nucl. Phys.}
  \textbf{\bibinfo{volume}{31}}, \bibinfo{pages}{7} (\bibinfo{year}{2007}),
  \eprint{hep-ph/0607226}.

\bibitem[{\citenamefont{Du et~al.}(2013{\natexlab{b}})\citenamefont{Du, Chen,
  Chen, and Zhu}}]{2013-Du-p14003-14003}
\bibinfo{author}{\bibfnamefont{M.-L.} \bibnamefont{Du}},
  \bibinfo{author}{\bibfnamefont{W.}~\bibnamefont{Chen}},
  \bibinfo{author}{\bibfnamefont{X.-L.} \bibnamefont{Chen}}, \bibnamefont{and}
  \bibinfo{author}{\bibfnamefont{S.-L.} \bibnamefont{Zhu}},
  \bibinfo{journal}{Phys.Rev.} \textbf{\bibinfo{volume}{D87}},
  \bibinfo{pages}{014003} (\bibinfo{year}{2013}{\natexlab{b}}).

\bibitem[{\citenamefont{Navarra et~al.}(2007)\citenamefont{Navarra, Nielsen,
  and Lee}}]{2007-Navarra-p166-172}
\bibinfo{author}{\bibfnamefont{F.~S.} \bibnamefont{Navarra}},
  \bibinfo{author}{\bibfnamefont{M.}~\bibnamefont{Nielsen}}, \bibnamefont{and}
  \bibinfo{author}{\bibfnamefont{S.~H.} \bibnamefont{Lee}},
  \bibinfo{journal}{Phys.Lett.} \textbf{\bibinfo{volume}{B649}},
  \bibinfo{pages}{166} (\bibinfo{year}{2007}), \eprint{hep-ph/0703071}.

\bibitem[{\citenamefont{Wang et~al.}(2011)\citenamefont{Wang, Xu, and
  Wang}}]{2011-Wang-p1049-1058}
\bibinfo{author}{\bibfnamefont{Z.-G.} \bibnamefont{Wang}},
  \bibinfo{author}{\bibfnamefont{Y.-M.} \bibnamefont{Xu}}, \bibnamefont{and}
  \bibinfo{author}{\bibfnamefont{H.-J.} \bibnamefont{Wang}},
  \bibinfo{journal}{Commun.Theor.Phys.} \textbf{\bibinfo{volume}{55}},
  \bibinfo{pages}{1049} (\bibinfo{year}{2011}).

\bibitem[{\citenamefont{Lipkin}(1973)}]{1973-Lipkin-p267-271}
\bibinfo{author}{\bibfnamefont{H.}~\bibnamefont{Lipkin}},
  \bibinfo{journal}{Phys.Lett.} \textbf{\bibinfo{volume}{B45}},
  \bibinfo{pages}{267} (\bibinfo{year}{1973}).

\bibitem[{\citenamefont{Ader et~al.}(1982)\citenamefont{Ader, Richard, and
  Taxil}}]{1982-Ader-p2370-2370}
\bibinfo{author}{\bibfnamefont{J.~P.}~\bibnamefont{Ader}},
  \bibinfo{author}{\bibfnamefont{J.~M.}~\bibnamefont{Richard}}, \bibnamefont{and}
  \bibinfo{author}{\bibfnamefont{P.}~\bibnamefont{Taxil}},
  \bibinfo{journal}{Phys.Rev.} \textbf{\bibinfo{volume}{D25}},
  \bibinfo{pages}{2370} (\bibinfo{year}{1982}).

\bibitem[{\citenamefont{Lipkin}(1986)}]{1986-Lipkin-p242-242}
\bibinfo{author}{\bibfnamefont{H.~J.} \bibnamefont{Lipkin}},
  \bibinfo{journal}{Phys.Lett.} \textbf{\bibinfo{volume}{B172}},
  \bibinfo{pages}{242} (\bibinfo{year}{1986}).


\bibitem[{\citenamefont{Richard}(1991)}]{1991-Richard-p254-257}
\bibinfo{author}{\bibfnamefont{J.~M.} \bibnamefont{Richard}},
  \bibinfo{journal}{Nucl.Phys.Proc.Suppl.} \textbf{\bibinfo{volume}{21}},
  \bibinfo{pages}{254} (\bibinfo{year}{1991}).

\bibitem[{\citenamefont{Bander and Subbaraman}(1994)}]{1994-Bander-p5478-5480}
\bibinfo{author}{\bibfnamefont{M.}~\bibnamefont{Bander}} \bibnamefont{and}
  \bibinfo{author}{\bibfnamefont{A.}~\bibnamefont{Subbaraman}},
  \bibinfo{journal}{Phys.Rev.} \textbf{\bibinfo{volume}{D50}},
  \bibinfo{pages}{5478} (\bibinfo{year}{1994}), \eprint{hep-ph/9407309}.

\bibitem[{\citenamefont{Moinester}(1996)}]{1996-Moinester-p349-362}
\bibinfo{author}{\bibfnamefont{M.~A.} \bibnamefont{Moinester}},
  \bibinfo{journal}{Z.Phys.} \textbf{\bibinfo{volume}{A355}},
  \bibinfo{pages}{349} (\bibinfo{year}{1996}), \eprint{hep-ph/9506405}.

\bibitem[{\citenamefont{Gelman and Nussinov}(2003)}]{2003-Gelman-p296-304}
\bibinfo{author}{\bibfnamefont{B.~A.} \bibnamefont{Gelman}} \bibnamefont{and}
  \bibinfo{author}{\bibfnamefont{S.}~\bibnamefont{Nussinov}},
  \bibinfo{journal}{Phys.Lett.} \textbf{\bibinfo{volume}{B551}},
  \bibinfo{pages}{296} (\bibinfo{year}{2003}), \eprint{hep-ph/0209095}.


\bibitem[{\citenamefont{Carames et~al.}(2011)\citenamefont{Carames, Valcarce,
  and Vijande}}]{2011-Carames-p291-295}
\bibinfo{author}{\bibfnamefont{T.}~\bibnamefont{Carames}},
  \bibinfo{author}{\bibfnamefont{A.}~\bibnamefont{Valcarce}}, \bibnamefont{and}
  \bibinfo{author}{\bibfnamefont{J.}~\bibnamefont{Vijande}},
  \bibinfo{journal}{Phys.Lett.} \textbf{\bibinfo{volume}{B699}},
  \bibinfo{pages}{291} (\bibinfo{year}{2011}).

\bibitem[{\citenamefont{Feng et~al.}(2013)\citenamefont{Feng, Guo, and
  Zou}}]{2013-Feng-p-}
\bibinfo{author}{\bibfnamefont{G.~Q.} \bibnamefont{Feng}},
  \bibinfo{author}{\bibfnamefont{X.~H.} \bibnamefont{Guo}}, \bibnamefont{and}
  \bibinfo{author}{\bibfnamefont{B.~S.} \bibnamefont{Zou}}
  (\bibinfo{year}{2013}), \eprint{hep-ph/1309.7813}.

\bibitem[{\citenamefont{Zhang and Huang}(2009)}]{2009-Zhang-p56004-56004}
\bibinfo{author}{\bibfnamefont{J.-R.} \bibnamefont{Zhang}} \bibnamefont{and}
  \bibinfo{author}{\bibfnamefont{M.-Q.} \bibnamefont{Huang}},
  \bibinfo{journal}{Phys. Rev.} \textbf{\bibinfo{volume}{D80}},
  \bibinfo{pages}{056004} (\bibinfo{year}{2009}).

\bibitem[{\citenamefont{Sun et~al.}(2012)\citenamefont{Sun, Liu, Nielsen, and
  Zhu}}]{2012-Sun-p94008-94008}
\bibinfo{author}{\bibfnamefont{Z.-F.} \bibnamefont{Sun}},
  \bibinfo{author}{\bibfnamefont{X.}~\bibnamefont{Liu}},
  \bibinfo{author}{\bibfnamefont{M.}~\bibnamefont{Nielsen}}, \bibnamefont{and}
  \bibinfo{author}{\bibfnamefont{S.-L.} \bibnamefont{Zhu}},
  \bibinfo{journal}{Phys.Rev.} \textbf{\bibinfo{volume}{D85}},
  \bibinfo{pages}{094008} (\bibinfo{year}{2012}).

\bibitem[{\citenamefont{Albuquerque et~al.}(2012)\citenamefont{Albuquerque,
  Liu, and Nielsen}}]{2012-Albuquerque-p492-498}
\bibinfo{author}{\bibfnamefont{R.~M.} \bibnamefont{Albuquerque}},
  \bibinfo{author}{\bibfnamefont{X.}~\bibnamefont{Liu}}, \bibnamefont{and}
  \bibinfo{author}{\bibfnamefont{M.}~\bibnamefont{Nielsen}},
  \bibinfo{journal}{Phys.Lett.} \textbf{\bibinfo{volume}{B718}},
  \bibinfo{pages}{492} (\bibinfo{year}{2012}).

\bibitem[{\citenamefont{Shifman et~al.}(1979)\citenamefont{Shifman, Vainshtein,
  and Zakharov}}]{1979-Shifman-p385-447}
\bibinfo{author}{\bibfnamefont{M.~A.} \bibnamefont{Shifman}},
  \bibinfo{author}{\bibfnamefont{A.~I.} \bibnamefont{Vainshtein}},
  \bibnamefont{and} \bibinfo{author}{\bibfnamefont{V.~I.}
  \bibnamefont{Zakharov}}, \bibinfo{journal}{Nucl. Phys.}
  \textbf{\bibinfo{volume}{B147}}, \bibinfo{pages}{385} (\bibinfo{year}{1979}).

\bibitem[{\citenamefont{Reinders et~al.}(1985)\citenamefont{Reinders,
  Rubinstein, and Yazaki}}]{1985-Reinders-p1-1}
\bibinfo{author}{\bibfnamefont{L.~J.} \bibnamefont{Reinders}},
  \bibinfo{author}{\bibfnamefont{H.}~\bibnamefont{Rubinstein}},
  \bibnamefont{and} \bibinfo{author}{\bibfnamefont{S.}~\bibnamefont{Yazaki}},
  \bibinfo{journal}{Phys. Rep.} \textbf{\bibinfo{volume}{127}},
  \bibinfo{pages}{1} (\bibinfo{year}{1985}).

\bibitem[{\citenamefont{Colangelo}(2000)}]{2000-Colangelo-p1495-1576}
\bibinfo{author}{\bibfnamefont{P.} \bibnamefont{Colangelo}},
  \bibnamefont{and} \bibinfo{author}{\bibfnamefont{A.}~\bibnamefont{Khodjamirian}},
  \bibinfo{journal}{Frontier Part. Phys.}
  \textbf{\bibinfo{volume}{3}}, \bibinfo{pages}{1495} (\bibinfo{year}{2000}), \eprint{hep-ph/0010175}.

\bibitem[{\citenamefont{Chen et~al.}(2013)\citenamefont{Chen, Steele, Du, and
  Zhu}}]{2013-Chen-p-}
\bibinfo{author}{\bibfnamefont{W.}~\bibnamefont{Chen}},
  \bibinfo{author}{\bibfnamefont{T. G.}~\bibnamefont{Steele}},
  \bibinfo{author}{\bibfnamefont{M.-L.} \bibnamefont{Du}}, \bibnamefont{and}
  \bibinfo{author}{\bibfnamefont{S.-L.} \bibnamefont{Zhu}},
  \bibinfo{journal}{Eur. Phys. J. C} \textbf{\bibinfo{volume}{74}},
  \bibinfo{pages}{2773} (\bibinfo{year}{2014}).

\bibitem[{\citenamefont{Braaten et~al.}(1992)\citenamefont{Braaten, Narison,
  and Pich}}]{1992-Braaten-p581-612}
\bibinfo{author}{\bibfnamefont{E.}~\bibnamefont{Braaten}},
  \bibinfo{author}{\bibfnamefont{S.}~\bibnamefont{Narison}}, \bibnamefont{and}
  \bibinfo{author}{\bibfnamefont{A.}~\bibnamefont{Pich}},
  \bibinfo{journal}{Nuclear Physics B} \textbf{\bibinfo{volume}{373}},
  \bibinfo{pages}{581} (\bibinfo{year}{1992}).

\bibitem[{\citenamefont{Narison}(2012)}]{2012-Narison-p259-263}
\bibinfo{author}{\bibfnamefont{S.}~\bibnamefont{Narison}},
  \bibinfo{journal}{Phys.Lett.} \textbf{\bibinfo{volume}{B707}},
  \bibinfo{pages}{259} (\bibinfo{year}{2012}).

\bibitem[{\citenamefont{Narison}(2010)}]{2010-Narison-p559-559}
\bibinfo{author}{\bibfnamefont{S.}~\bibnamefont{Narison}},
  \bibinfo{journal}{Phys.Lett.} \textbf{\bibinfo{volume}{B693}},
  \bibinfo{pages}{559} (\bibinfo{year}{2010}).

\bibitem[{\citenamefont{Kuhn et~al.}(2007)\citenamefont{Kuhn, Steinhauser, and
  Sturm}}]{2007-Kuhn-p192-215}
\bibinfo{author}{\bibfnamefont{J.~H.} \bibnamefont{Kuhn}},
  \bibinfo{author}{\bibfnamefont{M.}~\bibnamefont{Steinhauser}},
  \bibnamefont{and} \bibinfo{author}{\bibfnamefont{C.}~\bibnamefont{Sturm}},
  \bibinfo{journal}{Nucl.Phys.} \textbf{\bibinfo{volume}{B778}},
  \bibinfo{pages}{192} (\bibinfo{year}{2007}).

\bibitem[{\citenamefont{Chen et~al.}(2013{\natexlab{b}})\citenamefont{Chen,
  Steele, and Zhu}}]{2013-Chen-p-a}
\bibinfo{author}{\bibfnamefont{W.}~\bibnamefont{Chen}},
  \bibinfo{author}{\bibfnamefont{T.~G.}~\bibnamefont{Steele}}, \bibnamefont{and}
  \bibinfo{author}{\bibfnamefont{S.-L.} \bibnamefont{Zhu}},
  \bibinfo{journal}{J. Phys. G} \textbf{\bibinfo{volume}{41}},
  \bibinfo{pages}{025003} (\bibinfo{year}{2014}).

\end{thebibliography}

\appendix

\section{SPECTRAL DENSITIES}\label{sec:rhos}
In this Appendix, we list the spectral densities of the tetraquark
interpolating currents in Eqs.~\eqref{current1}-\eqref{current4}. At
leading order in $\alpha_s$, we calculate the spectral densities
including the perturbative terms, quark condensate $\qq$, gluon
condensate $\GGa$, quark-gluon mixed condensate $\qGqa$ and
dimension eight condensate $\qq\qGqa$:
\begin{eqnarray}
\rho(s)=\rho^{pert}(s)+\rho^{\qq}(s)+\rho^{\GGa}(s)+\rho^{\qGqb}(s)+\rho^{\qq^2}(s)+\rho^{\qq\qGqb}(s).
\end{eqnarray}

\subsection{The spectral densities for the $bc\bar q\bar q$ and $bc\bar s\bar s$ systems}
For the interpolating current $J_1$ with $J^P=0^+$:
{\allowdisplaybreaks
\begin{eqnarray}
\nonumber\rho_1^{pert}(s)&=&\dab\frac{(1-\alpha-\beta)^2(m_1^2\beta+m_2^2\alpha-\alpha\beta
s)^3(m_1^2\beta+m_2^2\alpha-3\alpha\beta
s-2m_1m_2)}{256\pi^6\alpha^3\beta^3}, \non
\rho_1^{\qq}(s)&=&-m_q\qq\dab\frac{(m_1^2\beta+m_2^2\alpha-\alpha\beta
s)(m_1^2\beta+m_2^2\alpha-2\alpha\beta
s-m_1m_2)}{8\pi^4\alpha\beta}, \non
\rho_{1a}^{\GGa}(s)&=&\GGb\dab\frac{(1-\alpha-\beta)^2}{1536\pi^6}\bigg[(2m_1^2\beta+2m_2^2\alpha-3\alpha\beta
s)\left(\frac{m_1^2}{\alpha^3}+\frac{m_2^2}{\beta^3}\right)- \non
&&\frac{m_1m_2}{\alpha\beta}\bigg(\frac{4m_1^2\beta+3m_2^2\alpha-3\alpha\beta
s}{\alpha^2}+\frac{3m_1^2\beta+4m_2^2\alpha-3\alpha\beta
s}{\beta^2}\bigg)\bigg], \non
\rho_{1b}^{\GGa}(s)&=&-\GGb\dab\frac{(m_1^2\beta+m_2^2\alpha-\alpha\beta
s)}{1024\pi^6}\times
\non&&\bigg[\frac{m_1^2\beta+m_2^2\alpha-2\alpha\beta
s-m_1m_2}{\alpha\beta}+\frac{(1-\alpha-\beta)^2(m_1^2\beta+m_2^2\alpha-2\alpha\beta
s-2m_1m_2)}{2\alpha^2\beta^2}\bigg], \non
\rho_1^{\qGqb}(s)&=&-\frac{m_q\qGqa\left[s-(m_1-m_2)^2\right]}{32\pi^4}\left[\left(1+\frac{m_1^2-m_2^2}{s}\right)^2-\frac{4m_1^2}{s}\right]^{1/2},
\\
\rho_1^{\qq^2}(s)&=&\frac{\qq^2\left[s-(m_1-m_2)^2\right]}{12\pi^2}\left[\left(1+\frac{m_1^2-m_2^2}{s}\right)^2-\frac{4m_1^2}{s}\right]^{1/2},
\non
\rho_1^{\qq\qGqb}(s)&=&-\frac{\qq\qGqb}{12\pi^2}\int_0^1d\alpha\Bigg\{\left[\frac{m_2^4-m_1m_2^3}{\alpha^2}-\frac{m_2^4(1-\alpha)+m_1^2m_2^2\alpha}{\alpha^2(1-\alpha)}\right]
\delta'\left[s-\frac{m_1^2\alpha+m_2^2(1-\alpha)}{\alpha(1-\alpha)}\right]-
\non&&\frac{m_1^2\alpha+m_2^2(1-\alpha)}{\alpha(1-\alpha)}\delta\left[s-\frac{m_1^2\alpha+m_2^2(1-\alpha)}{\alpha(1-\alpha)}\right]-H\left[s-\frac{m_1^2\alpha+m_2^2(1-\alpha)}{\alpha(1-\alpha)}\right]\Bigg\}.
\end{eqnarray}
} where
$\alpha_{min}=\frac{1}{2}\left\{1+\frac{m_1^2-m_2^2}{s}-\left[\left(1+\frac{m_1^2-m_2^2}{s}\right)^2-\frac{4m_1^2}{s}\right]^{1/2}\right\}$,
$\alpha_{max}=\frac{1}{2}\left\{1+\frac{m_1^2-m_2^2}{s}+\left[\left(1+\frac{m_1^2-m_2^2}{s}\right)^2-\frac{4m_1^2}{s}\right]^{1/2}\right\}$,
$\beta_{min}=\frac{\alpha m_2^2}{\alpha s-m_1^2}$,
$\beta_{max}=1-\alpha$. $m_1$ and $m_2$ are the heavy quark masses.
$H(\alpha)$ is the Heaviside step function.

For the interpolating current $J_3$ with $J^P=0^+$:
\begin{eqnarray}
\nonumber\rho_3^{pert}(s)&=&\frac{1}{2}\rho_1^{pert}(s),
\rho_3^{\qq}(s)=\frac{1}{2}\rho_1^{\qq}(s),
\rho_{3a}^{\GGa}(s)=\frac{1}{2}\rho_{1a}^{\GGa}(s),
\rho_{3b}^{\GGa}(s)=-\rho_{1b}^{\GGa}(s),
\\
\rho_3^{\qGqb}(s)&=&\frac{1}{2}\rho_1^{\qGqb}(s),
\rho_3^{\qq^2}(s)=\frac{1}{2}\rho_1^{\qq^2}(s),
\rho_3^{\qq\qGqb}(s)=\frac{1}{2}\rho_1^{\qq\qGqb}(s).
\end{eqnarray}

For the interpolating current $J_2$ with $J^P=0^+$:
{\allowdisplaybreaks
\begin{eqnarray}
\nonumber\rho_2^{pert}(s)&=&\dab\frac{(1-\alpha-\beta)^2(m_1^2\beta+m_2^2\alpha-\alpha\beta
s)^3(m_1^2\beta+m_2^2\alpha-3\alpha\beta
s-4m_1m_2)}{64\pi^6\alpha^3\beta^3}, \non
\rho_2^{\qq}(s)&=&-m_q\qq\dab\frac{3(m_1^2\beta+m_2^2\alpha-\alpha\beta
s)(m_1^2\beta+m_2^2\alpha-2\alpha\beta
s-2m_1m_2)}{2\pi^4\alpha\beta}, \non
\rho_{2a}^{\GGa}(s)&=&\GGb\dab\frac{(1-\alpha-\beta)^2}{384\pi^6}\bigg[(2m_1^2\beta+2m_2^2\alpha-3\alpha\beta
s)\left(\frac{m_1^2}{\alpha^3}+\frac{m_2^2}{\beta^3}\right)- \non
&&\frac{2m_1m_2}{\alpha\beta}\bigg(\frac{4m_1^2\beta+3m_2^2\alpha-3\alpha\beta
s}{\alpha^2}+\frac{3m_1^2\beta+4m_2^2\alpha-3\alpha\beta
s}{\beta^2}\bigg)\bigg], \non
\rho_{2b}^{\GGa}(s)&=&\GGb\dab\frac{(m_1^2\beta+m_2^2\alpha-\alpha\beta
s)}{256\pi^6}\times
\non&&\bigg[\frac{m_1^2\beta+m_2^2\alpha-2\alpha\beta
s-2m_1m_2}{\alpha\beta}+\frac{(1-\alpha-\beta)^2(m_1^2\beta+m_2^2\alpha-2\alpha\beta
s)}{2\alpha^2\beta^2}\bigg], \non
\rho_2^{\qGqb}(s)&=&-\frac{m_q\qGqa\left[s-(m_1-m_2)^2+2m_1m_2\right]}{4\pi^4}\left[\left(1+\frac{m_1^2-m_2^2}{s}\right)^2-\frac{4m_1^2}{s}\right]^{1/2},
\\
\rho_2^{\qq^2}(s)&=&\frac{2\qq^2\left[s-(m_1-m_2)^2+2m_1m_2\right]}{3\pi^2}\left[\left(1+\frac{m_1^2-m_2^2}{s}\right)^2-\frac{4m_1^2}{s}\right]^{1/2},
\non
\rho_2^{\qq\qGqb}(s)&=&-\frac{2\qq\qGqb}{3\pi^2}\int_0^1d\alpha\Bigg\{\left[\frac{m_2^4-2m_1m_2^3}{\alpha^2}-\frac{m_2^4(1-\alpha)+m_1^2m_2^2\alpha}{\alpha^2(1-\alpha)}\right]
\delta'\left[s-\frac{m_1^2\alpha+m_2^2(1-\alpha)}{\alpha(1-\alpha)}\right]-
\non&&\frac{m_1^2\alpha+m_2^2(1-\alpha)}{\alpha(1-\alpha)}\delta\left[s-\frac{m_1^2\alpha+m_2^2(1-\alpha)}{\alpha(1-\alpha)}\right]-H\left[s-\frac{m_1^2\alpha+m_2^2(1-\alpha)}{\alpha(1-\alpha)}\right]\Bigg\}.
\end{eqnarray}
}

For the interpolating current $J_4$ with $J^P=0^+$:
\begin{eqnarray}
\nonumber\rho_4^{pert}(s)&=&\frac{1}{2}\rho_2^{pert}(s),
\rho_4^{\qq}(s)=\frac{1}{2}\rho_2^{\qq}(s),
\rho_{4a}^{\GGa}(s)=\frac{1}{2}\rho_{2a}^{\GGa}(s),
\rho_{4b}^{\GGa}(s)=-\rho_{2b}^{\GGa}(s),
\\
\rho_4^{\qGqb}(s)&=&\frac{1}{2}\rho_2^{\qGqb}(s),
\rho_4^{\qq^2}(s)=\frac{1}{2}\rho_2^{\qq^2}(s),
\rho_4^{\qq\qGqb}(s)=\frac{1}{2}\rho_2^{\qq\qGqb}(s).
\end{eqnarray}

For the interpolating current $J_1$ with $J^P=1^+$:
{\allowdisplaybreaks
\begin{eqnarray}
\nonumber\rho_1^{pert}(s)&=&\dab\frac{(1-\alpha-\beta)^2(m_1^2\beta+m_2^2\alpha-\alpha\beta
s)^3}{1536\pi^6\alpha^3\beta^3}\times
\non&&\Big[6(m_1^2\beta+m_2^2\alpha-3\alpha\beta
s-2m_1m_2)-(1-\alpha-\beta)(3m_1^2\beta+3m_2^2\alpha-7\alpha\beta
s-4m_1m_2)\Big], \non
\rho_1^{\qq}(s)&=&-\frac{m_q\qq}{16\pi^4}\dab\frac{(m_1^2\beta+m_2^2\alpha-\alpha\beta
s)}{\alpha\beta}\times
\non&&\Big[2(m_1^2\beta+m_2^2\alpha-2\alpha\beta
s-m_1m_2)-(1-\alpha-\beta)(3m_1^2\beta+3m_2^2\alpha-5\alpha\beta
s-2m_1m_2)\Big], \non
\rho_{1a}^{\GGa}(s)&=&\GGb\dab\frac{(1-\alpha-\beta)^2}{4608\pi^6}\Bigg\{
\non&&\left(\frac{m_1^2}{\alpha^3}+\frac{m_2^2}{\beta^3}\right)\Big[3(2m_1^2\beta+2m_2^2\alpha-3\alpha\beta
s)-(1-\alpha-\beta)(3m_1^2\beta+3m_2^2\alpha-4\alpha\beta s)\Big]-
\non
&&\frac{m_1m_2(2+\alpha+\beta)}{\alpha\beta}\bigg(\frac{4m_1^2\beta+3m_2^2\alpha-3\alpha\beta
s}{\alpha^2}+\frac{3m_1^2\beta+4m_2^2\alpha-3\alpha\beta
s}{\beta^2}\bigg)\Bigg\}, \non
\rho_{1b}^{\GGa}(s)&=&\GGb\dab\frac{(m_1^2\beta+m_2^2\alpha-\alpha\beta
s)}{12288\pi^6}\Bigg\{(1-\alpha-\beta)^2\times
\non&&\bigg[\frac{(1-\alpha-\beta)(3m_1^2\beta+3m_2^2\alpha-5\alpha\beta
s-4m_1m_2)}{\alpha^2\beta^2}-\frac{6(m_1^2\beta+m_2^2\alpha-2\alpha\beta
s-2m_1m_2)}{\alpha^2\beta^2}\bigg]+
\non&&\bigg[\frac{2(1-\alpha-\beta)(3m_1^2\beta+3m_2^2\alpha-5\alpha\beta
s-2m_1m_2)}{\alpha\beta}+\frac{4(m_1^2\beta+m_2^2\alpha-2\alpha\beta
s-m_1m_2)}{\alpha\beta}\bigg]\Bigg\}, \non
\rho_1^{\qGqb}(s)&=&-\frac{m_q\qGqa\left[s-(m_1-m_2)^2\right]}{32\pi^4}\left[\left(1+\frac{m_1^2-m_2^2}{s}\right)^2-\frac{4m_1^2}{s}\right]^{1/2},
\\
\rho_1^{\qq^2}(s)&=&\frac{\qq^2\left[s-(m_1-m_2)^2\right]}{12\pi^2}\left[\left(1+\frac{m_1^2-m_2^2}{s}\right)^2-\frac{4m_1^2}{s}\right]^{1/2},
\non
\rho_1^{\qq\qGqb}(s)&=&-\frac{\qq\qGqb}{12\pi^2}\int_0^1d\alpha\Bigg\{\left[\frac{m_2^4-m_1m_2^3}{\alpha^2}-\frac{m_2^4(1-\alpha)+m_1^2m_2^2\alpha}{\alpha^2(1-\alpha)}\right]
\delta'\left[s-\frac{m_1^2\alpha+m_2^2(1-\alpha)}{\alpha(1-\alpha)}\right]-
\non&&\frac{m_1^2\alpha+m_2^2(1-\alpha)}{\alpha(1-\alpha)}\delta\left[s-\frac{m_1^2\alpha+m_2^2(1-\alpha)}{\alpha(1-\alpha)}\right]-H\left[s-\frac{m_1^2\alpha+m_2^2(1-\alpha)}{\alpha(1-\alpha)}\right]\Bigg\}.
\end{eqnarray}
}

For the interpolating current $J_3$ with $J^P=1^+$:
\begin{eqnarray}
\nonumber\rho_3^{pert}(s)&=&\frac{1}{2}\rho_1^{pert}(s),
\rho_3^{\qq}(s)=\frac{1}{2}\rho_1^{\qq}(s),
\rho_{3a}^{\GGa}(s)=\frac{1}{2}\rho_{1a}^{\GGa}(s),
\rho_{3b}^{\GGa}(s)=-\rho_{1b}^{\GGa}(s),
\\
\rho_3^{\qGqb}(s)&=&\frac{1}{2}\rho_1^{\qGqb}(s),
\rho_3^{\qq^2}(s)=\frac{1}{2}\rho_1^{\qq^2}(s),
\rho_3^{\qq\qGqb}(s)=\frac{1}{2}\rho_1^{\qq\qGqb}(s).
\end{eqnarray}

For the interpolating current $J_2$ with $J^P=1^+$:
{\allowdisplaybreaks
\begin{eqnarray}
\nonumber\rho_2^{pert}(s)&=&\dab\frac{(1-\alpha-\beta)^2(m_1^2\beta+m_2^2\alpha-\alpha\beta
s)^3(m_1^2\beta+m_2^2\alpha-5\alpha\beta
s-4m_1m_2)}{512\pi^6\alpha^3\beta^3}, \non
\rho_2^{\qq}(s)&=&-m_q\qq\dab\frac{(m_1^2\beta+m_2^2\alpha-\alpha\beta
s)(m_1^2\beta+m_2^2\alpha-3\alpha\beta
s-2m_1m_2)}{16\pi^4\alpha\beta}, \non
\rho_{2a}^{\GGa}(s)&=&\GGb\dab\frac{(1-\alpha-\beta)^2}{1536\pi^6}\bigg[(m_1^2\beta+m_2^2\alpha-2\alpha\beta
s)\left(\frac{m_1^2}{\alpha^3}+\frac{m_2^2}{\beta^3}\right)- \non
&&\frac{m_1m_2}{\alpha\beta}\bigg(\frac{4m_1^2\beta+3m_2^2\alpha-3\alpha\beta
s}{\alpha^2}+\frac{3m_1^2\beta+4m_2^2\alpha-3\alpha\beta
s}{\beta^2}\bigg)\bigg], \non
\rho_{2b}^{\GGa}(s)&=&-\GGb\dab\frac{(m_1^2\beta+m_2^2\alpha-\alpha\beta
s)}{12288\pi^6}\times
\non&&\bigg[\frac{6(m_1^2\beta+m_2^2\alpha-3\alpha\beta
s-2m_1m_2)}{\alpha\beta}-\frac{(1-\alpha-\beta)^2(3m_1^2\beta+3m_2^2\alpha-5\alpha\beta
s)}{\alpha^2\beta^2}\bigg], \non
\rho_2^{\qGqb}(s)&=&-\frac{m_q\qGqa\left[2s^2-(m_1^2-6m_1m_2+m_2^2)s-(m_1^2-m_2^2)^2\right]}{96\pi^4
s}\left[\left(1+\frac{m_1^2-m_2^2}{s}\right)^2-\frac{4m_1^2}{s}\right]^{1/2},
\\
\rho_2^{\qq^2}(s)&=&\frac{\qq^2\left[2s^2-(m_1^2-6m_1m_2+m_2^2)s-(m_1^2-m_2^2)^2\right]}{36\pi^2
s}\left[\left(1+\frac{m_1^2-m_2^2}{s}\right)^2-\frac{4m_1^2}{s}\right]^{1/2},
\non
\rho_2^{\qq\qGqb}(s)&=&-\frac{\qq\qGqb}{12\pi^2}\int_0^1d\alpha\Bigg\{\left[\frac{m_2^4-m_1m_2^3}{\alpha^2}-\frac{m_2^4(1-\alpha)+m_1^2m_2^2\alpha}{\alpha^2(1-\alpha)}\right]
\delta'\left[s-\frac{m_1^2\alpha+m_2^2(1-\alpha)}{\alpha(1-\alpha)}\right]-
\non&&\frac{m_1^2\alpha+m_2^2(1-\alpha)}{\alpha(1-\alpha)}\delta\left[s-\frac{m_1^2\alpha+m_2^2(1-\alpha)}{\alpha(1-\alpha)}\right]-\alpha*H\left[s-\frac{m_1^2\alpha+m_2^2(1-\alpha)}{\alpha(1-\alpha)}\right]\Bigg\}.
\end{eqnarray}
}

For the interpolating current $J_4$ with $J^P=1^+$:
\begin{eqnarray}
\nonumber\rho_4^{pert}(s)&=&\frac{1}{2}\rho_2^{pert}(s),
\rho_4^{\qq}(s)=\frac{1}{2}\rho_2^{\qq}(s),
\rho_{4a}^{\GGa}(s)=\frac{1}{2}\rho_{2a}^{\GGa}(s),
\rho_{4b}^{\GGa}(s)=-\rho_{2b}^{\GGa}(s),
\\
\rho_4^{\qGqb}(s)&=&\frac{1}{2}\rho_2^{\qGqb}(s),
\rho_4^{\qq^2}(s)=\frac{1}{2}\rho_2^{\qq^2}(s),
\rho_4^{\qq\qGqb}(s)=\frac{1}{2}\rho_2^{\qq\qGqb}(s).
\end{eqnarray}

\subsection{The spectral densities for the $qc\bar q\bar b$ and $sc\bar s\bar b$ systems}
For the interpolating current $J_1$ with $J^P=0^+$:
{\allowdisplaybreaks
\begin{eqnarray}
\nonumber\rho_1^{pert}(s)&=&\dab\frac{(1-\alpha-\beta)^2(m_1^2\beta+m_2^2\alpha-\alpha\beta
s)^2}{256\pi^6}\times
\non&&\Bigg[\frac{(m_1^2\beta+m_2^2\alpha-3\alpha\beta
s)(m_1^2\beta+m_2^2\alpha-\alpha\beta
s)}{\alpha^3\beta^3}-\left(\frac{m_1}{\alpha}+\frac{m_2}{\beta}\right)\frac{2m_q(2m_1^2\beta+2m_2^2\alpha-5\alpha\beta
s)}{\alpha^2\beta^2}\Bigg], \non
\rho_1^{\qq}(s)&=&\qq\dab\frac{(m_1^2\beta+m_2^2\alpha-\alpha\beta
s)}{8\pi^4\alpha\beta}\times
\non&&\Bigg[m_q(m_1^2\beta+m_2^2\alpha-2\alpha\beta
s+2m_1m_2)-\left(\frac{m_1}{\alpha}+\frac{m_2}{\beta}\right)(1-\alpha-\beta)(m_1^2\beta+m_2^2\alpha-2\alpha\beta
s)\Bigg], \non
\rho_{1a}^{\GGa}(s)&=&\GGb\dab\frac{(1-\alpha-\beta)^2(2m_1^2\beta+2m_2^2\alpha-3\alpha\beta
s)}{1536\pi^6}\left(\frac{m_1^2}{\alpha^3}+\frac{m_2^2}{\beta^3}\right),
\non
\rho_{1b}^{\GGa}(s)&=&-\GGb\dab\frac{(1-\alpha-\beta)(m_1^2\beta+m_2^2\alpha-2\alpha\beta
s)(m_1^2\beta+m_2^2\alpha-\alpha\beta
s)}{1024\pi^6\alpha\beta}\left(\frac{1}{\alpha}+\frac{1}{\beta}\right),
\non
\rho_{1a}^{\qGqb}(s)&=&\qGqa\dab\frac{(2m_1^2\beta+2m_2^2\alpha-3\alpha\beta
s)}{32\pi^4}\left(\frac{m_1}{\alpha}+\frac{m_2}{\beta}\right), \non
\rho_{1b}^{\qGqb}(s)&=&\qGqa\dab
\non&&\Bigg[\frac{(1-\alpha-\beta)(2m_1^2\beta+2m_2^2\alpha-3\alpha\beta
s)}{64\pi^4}\left(\frac{m_1}{\alpha^2}+\frac{m_2}{\beta^2}\right)-\frac{m_qm_1m_2}{64\pi^4}\left(\frac{1}{\alpha}+\frac{1}{\beta}\right)\Bigg],
\non
\rho_{1c}^{\qGqb}(s)&=&-\frac{m_qm_1m_2\qGqa}{16\pi^4}\left[\left(1+\frac{m_1^2-m_2^2}{s}\right)^2-\frac{4m_1^2}{s}\right]^{1/2},
\non
\rho_1^{\qq^2}(s)&=&\frac{\qq^2}{12\pi^2}\left[\left(1+\frac{m_1^2-m_2^2}{s}\right)^2-\frac{4m_1^2}{s}\right]^{1/2}
\Bigg\{2m_1m_2-\frac{m_qm_1(m_1^2-m_2^2-s)+m_qm_2(m_1^2-m_2^2+s)}{s}
\non&&+\frac{m_qm_1}{s[(m_1^2-m_2^2-s)^2-4m_2^2s]}\left[m_1^6-m_2^6+m_2^4s-m_1^4(3m_2^2+2s)+m_1^2(3m_2^4+m_2^2s+s^2)\right]
\non&&+\frac{m_qm_2}{s[(m_1^2-m_2^2-s)^2-4m_2^2s]}\left[m_1^6-m_1^4(3m_2^2+s)+m_1^2(3m_2^4-m_2^2s)-(m_2^3-m_2s)^2\right]\Bigg\},
\non
\rho_{1a}^{\qq\qGqb}(s)&=&\frac{\qq\qGqb}{12\pi^2}\int_0^1d\alpha\frac{m_1m_2^3}{\alpha^2}\delta'\left[s-\frac{m_1^2\alpha+m_2^2(1-\alpha)}{\alpha(1-\alpha)}\right],
\\
\rho_{1b}^{\qq\qGqb}(s)&=&\frac{\qq\qGqb}{48\pi^2}\int_0^1d\alpha\frac{m_1m_2}{\alpha(1-\alpha)}\delta\left[s-\frac{m_1^2\alpha+m_2^2(1-\alpha)}{\alpha(1-\alpha)}\right]. \label{eq1}
\end{eqnarray}
}

For the interpolating current $J_3$ with $J^P=0^+$:
\begin{eqnarray}
\nonumber\rho_3^{pert}(s)&=&\frac{1}{2}\rho_1^{pert}(s),
\rho_3^{\qq}(s)=\frac{1}{2}\rho_1^{\qq}(s),
\rho_{3a}^{\GGa}(s)=\frac{1}{2}\rho_{1a}^{\GGa}(s),
\rho_{3b}^{\GGa}(s)=-\rho_{1b}^{\GGa}(s), \non
\rho_{3a}^{\qGqb}(s)&=&\frac{1}{2}\rho_{1a}^{\qGqb}(s),
\rho_{3b}^{\qGqb}(s)=-\rho_{1b}^{\qGqb}(s),
\rho_{3c}^{\qGqb}(s)=\frac{1}{2}\rho_{1c}^{\qGqb}(s),
\rho_3^{\qq^2}(s)=\frac{1}{2}\rho_1^{\qq^2}(s),
\\ \rho_{3a}^{\qq\qGqb}(s)&=&\frac{1}{2}\rho_{1a}^{\qq\qGqb}(s), \rho_{3b}^{\qq\qGqb}(s)=-\rho_{1b}^{\qq\qGqb}(s).
\end{eqnarray}

For the interpolating current $J_2$ with $J^P=0^+$:
{\allowdisplaybreaks
\begin{eqnarray}
\nonumber\rho_2^{pert}(s)&=&\dab\frac{(1-\alpha-\beta)^2(m_1^2\beta+m_2^2\alpha-\alpha\beta
s)^2}{64\pi^6}\times
\non&&\Bigg[\frac{(m_1^2\beta+m_2^2\alpha-3\alpha\beta
s)(m_1^2\beta+m_2^2\alpha-\alpha\beta
s)}{\alpha^3\beta^3}-\left(\frac{m_1}{\alpha}+\frac{m_2}{\beta}\right)\frac{4m_q(2m_1^2\beta+2m_2^2\alpha-5\alpha\beta
s)}{\alpha^2\beta^2}\Bigg], \non
\rho_2^{\qq}(s)&=&\qq\dab\frac{(m_1^2\beta+m_2^2\alpha-\alpha\beta
s)}{2\pi^4\alpha\beta}\times
\non&&\Bigg[m_q(m_1^2\beta+m_2^2\alpha-2\alpha\beta
s+8m_1m_2)-\left(\frac{m_1}{\alpha}+\frac{m_2}{\beta}\right)(1-\alpha-\beta)(2m_1^2\beta+2m_2^2\alpha-4\alpha\beta
s)\Bigg], \non
\rho_{2a}^{\GGa}(s)&=&\GGb\dab\frac{(1-\alpha-\beta)^2(2m_1^2\beta+2m_2^2\alpha-3\alpha\beta
s)}{384\pi^6}\left(\frac{m_1^2}{\alpha^3}+\frac{m_2^2}{\beta^3}\right),
\non
\rho_{2b}^{\GGa}(s)&=&\GGb\dab\frac{(1-\alpha-\beta)(m_1^2\beta+m_2^2\alpha-2\alpha\beta
s)(m_1^2\beta+m_2^2\alpha-\alpha\beta
s)}{256\pi^6\alpha\beta}\left(\frac{1}{\alpha}+\frac{1}{\beta}\right),
\non
\rho_{2a}^{\qGqb}(s)&=&\qGqa\dab\frac{(2m_1^2\beta+2m_2^2\alpha-3\alpha\beta
s)}{4\pi^4}\left(\frac{m_1}{\alpha}+\frac{m_2}{\beta}\right), \non
\rho_{2b}^{\qGqb}(s)&=&-\frac{m_qm_1m_2\qGqa}{\pi^4}\left[\left(1+\frac{m_1^2-m_2^2}{s}\right)^2-\frac{4m_1^2}{s}\right]^{1/2},
\non
\rho_2^{\qq^2}(s)&=&\frac{2\qq^2}{3\pi^2}\left[\left(1+\frac{m_1^2-m_2^2}{s}\right)^2-\frac{4m_1^2}{s}\right]^{1/2}
\Bigg\{4m_1m_2-\frac{m_qm_1(m_1^2-m_2^2-s)+m_qm_2(m_1^2-m_2^2+s)}{s}
\non&&+\frac{m_qm_1}{s[(m_1^2-m_2^2-s)^2-4m_2^2s]}\left[m_1^6-m_2^6+m_2^4s-m_1^4(3m_2^2+2s)+m_1^2(3m_2^4+m_2^2s+s^2)\right]
\non&&+\frac{m_qm_2}{s[(m_1^2-m_2^2-s)^2-4m_2^2s]}\left[m_1^6-m_1^4(3m_2^2+s)+m_1^2(3m_2^4-m_2^2s)-(m_2^3-m_2s)^2\right]\Bigg\},
\\
\rho_{2}^{\qq\qGqb}(s)&=&\frac{4\qq\qGqb}{3\pi^2}\int_0^1d\alpha\frac{m_1m_2^3}{\alpha^2}\delta'\left[s-\frac{m_1^2\alpha+m_2^2(1-\alpha)}{\alpha(1-\alpha)}\right],
\end{eqnarray}
}

For the interpolating current $J_4$ with $J^P=0^+$:
\begin{eqnarray}
\nonumber\rho_4^{pert}(s)&=&\frac{1}{2}\rho_2^{pert}(s),
\rho_4^{\qq}(s)=\frac{1}{2}\rho_2^{\qq}(s),
\rho_{4a}^{\GGa}(s)=\frac{1}{2}\rho_{2a}^{\GGa}(s),
\rho_{4b}^{\GGa}(s)=-\rho_{2b}^{\GGa}(s),
\\
\rho_{4a}^{\qGqb}(s)&=&\frac{1}{2}\rho_{2a}^{\qGqb}(s),
\rho_{4b}^{\qGqb}(s)=\frac{1}{2}\rho_{2b}^{\qGqb}(s),
\rho_3^{\qq^2}(s)=\frac{1}{2}\rho_1^{\qq^2}(s),
\rho_{4}^{\qq\qGqb}(s)=\frac{1}{2}\rho_{2}^{\qq\qGqb}(s).
\end{eqnarray}

For the interpolating current $J_1$ with $J^P=1^+$:
{\allowdisplaybreaks
\begin{eqnarray}
\nonumber\rho_1^{pert}(s)&=&\dab\frac{(1-\alpha-\beta)^2(m_1^2\beta+m_2^2\alpha-\alpha\beta
s)^2}{512\pi^6}\Bigg[
\frac{4m_qm_2(m_1^2\beta+m_2^2\alpha-\alpha\beta
s)}{\alpha^2\beta^3}+
\non&&\frac{(m_1^2\beta+m_2^2\alpha-5\alpha\beta
s)(m_1^2\beta+m_2^2\alpha-\alpha\beta
s)}{\alpha^3\beta^3}-\left(\frac{m_1}{\alpha}+\frac{m_2}{\beta}\right)\frac{4m_q(2m_1^2\beta+2m_2^2\alpha-5\alpha\beta
s)}{\alpha^2\beta^2}\Bigg], \non
\rho_1^{\qq}(s)&=&\qq\dab\frac{(m_1^2\beta+m_2^2\alpha-\alpha\beta
s)}{16\pi^4}\Bigg[
\frac{m_2(1-\alpha-\beta)(m_1^2\beta+m_2^2\alpha-\alpha\beta
s)}{\alpha\beta^2}+
\non&&\frac{m_q(m_1^2\beta+m_2^2\alpha-3\alpha\beta
s+4m_1m_2)}{\alpha\beta}-\left(\frac{m_1}{\alpha}+\frac{m_2}{\beta}\right)\frac{2(1-\alpha-\beta)(m_1^2\beta+m_2^2\alpha-2\alpha\beta
s)}{\alpha\beta}\Bigg], \non
\rho_{1a}^{\GGa}(s)&=&\GGb\dab\frac{(1-\alpha-\beta)^2(m_1^2\beta+m_2^2\alpha-2\alpha\beta
s)}{1536\pi^6}\left(\frac{m_1^2}{\alpha^3}+\frac{m_2^2}{\beta^3}\right),
\non \rho_{1b}^{\GGa}(s)&=&\GGb\dab
\non&&\frac{(1-\alpha-\beta)(m_1^2\beta+m_2^2\alpha-\alpha\beta
s)}{6144\pi^6\alpha\beta}\left(\frac{3m_1^2\beta+3m_2^2\alpha-5\alpha\beta
s}{\alpha}-\frac{3m_1^2\beta+3m_2^2\alpha-9\alpha\beta
s}{\beta}\right), \non
\rho_{1a}^{\qGqb}(s)&=&\qGqa\dab\left[\frac{m_1(2m_1^2\beta+2m_2^2\alpha-3\alpha\beta
s)}{32\pi^4 \alpha}+\frac{m_2(m_1^2\beta+m_2^2\alpha-2\alpha\beta
s)}{32\pi^4 \beta}\right], \non
\rho_{1b}^{\qGqb}(s)&=&\qGqa\dab\Bigg[\frac{m_2(1-\alpha-\beta)(m_1^2\beta+m_2^2\alpha-2\alpha\beta
s)}{64\pi^4\beta^2}-\frac{m_qm_1m_2}{64\pi^4\beta}\Bigg], \non
\rho_{1c}^{\qGqb}(s)&=&-\frac{m_qm_1m_2\qGqa}{16\pi^4}\left[\left(1+\frac{m_1^2-m_2^2}{s}\right)^2-\frac{4m_1^2}{s}\right]^{1/2},
\non
\rho_1^{\qq^2}(s)&=&\frac{\qq^2}{24\pi^2}\left[\left(1+\frac{m_1^2-m_2^2}{s}\right)^2-\frac{4m_1^2}{s}\right]^{1/2}
\Bigg\{4m_1m_2-\frac{2m_qm_1(m_1^2-m_2^2-s)+m_qm_2(m_1^2-m_2^2+s)}{s}
\non&&+\frac{2m_qm_1}{s[(m_1^2-m_2^2-s)^2-4m_2^2s]}\left[m_1^6-m_2^6+m_2^4s-m_1^4(3m_2^2+2s)+m_1^2(3m_2^4+m_2^2s+s^2)\right]
\non&&+\frac{2m_qm_2}{s[(m_1^2-m_2^2-s)^2-4m_2^2s]}\left[m_1^6-m_1^4(3m_2^2+s)+m_1^2(3m_2^4-m_2^2s)-(m_2^3-m_2s)^2\right]\Bigg\},
\non
\rho_{1a}^{\qq\qGqb}(s)&=&\frac{\qq\qGqb}{12\pi^2}\int_0^1d\alpha\frac{m_1m_2^3}{\alpha^2}\delta'\left[s-\frac{m_1^2\alpha+m_2^2(1-\alpha)}{\alpha(1-\alpha)}\right],
\\
\rho_{1b}^{\qq\qGqb}(s)&=&\frac{\qq\qGqb}{48\pi^2}\int_0^1d\alpha\frac{m_1m_2}{\alpha}\delta\left[s-\frac{m_1^2\alpha+m_2^2(1-\alpha)}{\alpha(1-\alpha)}\right].
\end{eqnarray}
}

For the interpolating current $J_3$ with $J^P=1^+$:
\begin{eqnarray}
\nonumber\rho_3^{pert}(s)&=&\frac{1}{2}\rho_1^{pert}(s),
\rho_3^{\qq}(s)=\frac{1}{2}\rho_1^{\qq}(s),
\rho_{3a}^{\GGa}(s)=\frac{1}{2}\rho_{1a}^{\GGa}(s),
\rho_{3b}^{\GGa}(s)=-\rho_{1b}^{\GGa}(s), \non
\rho_{3a}^{\qGqb}(s)&=&\frac{1}{2}\rho_{1a}^{\qGqb}(s),
\rho_{3b}^{\qGqb}(s)=-\rho_{1b}^{\qGqb}(s),
\rho_{3c}^{\qGqb}(s)=\frac{1}{2}\rho_{1c}^{\qGqb}(s),
\rho_3^{\qq^2}(s)=\frac{1}{2}\rho_1^{\qq^2}(s),
\\ \rho_{3a}^{\qq\qGqb}(s)&=&\frac{1}{2}\rho_{1a}^{\qq\qGqb}(s), \rho_{3b}^{\qq\qGqb}(s)=-\rho_{1b}^{\qq\qGqb}(s).
\end{eqnarray}

For the interpolating current $J_2$ with $J^P=1^+$:
{\allowdisplaybreaks
\begin{eqnarray}
\nonumber\rho_2^{pert}(s)&=&\dab\frac{(1-\alpha-\beta)^2(m_1^2\beta+m_2^2\alpha-\alpha\beta
s)^2}{512\pi^6}\Bigg[
\frac{4m_qm_1(m_1^2\beta+m_2^2\alpha-\alpha\beta
s)}{\alpha^3\beta^2}+
\non&&\frac{(m_1^2\beta+m_2^2\alpha-5\alpha\beta
s)(m_1^2\beta+m_2^2\alpha-\alpha\beta
s)}{\alpha^3\beta^3}-\left(\frac{m_1}{\alpha}+\frac{m_2}{\beta}\right)\frac{4m_q(2m_1^2\beta+2m_2^2\alpha-5\alpha\beta
s)}{\alpha^2\beta^2}\Bigg], \non
\rho_2^{\qq}(s)&=&\qq\dab\frac{(m_1^2\beta+m_2^2\alpha-\alpha\beta
s)}{16\pi^4}\Bigg[
\frac{m_1(1-\alpha-\beta)(m_1^2\beta+m_2^2\alpha-\alpha\beta
s)}{\alpha^2\beta}+
\non&&\frac{m_q(m_1^2\beta+m_2^2\alpha-3\alpha\beta
s+4m_1m_2)}{\alpha\beta}-\left(\frac{m_1}{\alpha}+\frac{m_2}{\beta}\right)\frac{2(1-\alpha-\beta)(m_1^2\beta+m_2^2\alpha-2\alpha\beta
s)}{\alpha\beta}\Bigg], \non
\rho_{2a}^{\GGa}(s)&=&\GGb\dab\frac{(1-\alpha-\beta)^2(m_1^2\beta+m_2^2\alpha-2\alpha\beta
s)}{1536\pi^6}\left(\frac{m_1^2}{\alpha^3}+\frac{m_2^2}{\beta^3}\right),
\non \rho_{2b}^{\GGa}(s)&=&\GGb\dab
\non&&\frac{(1-\alpha-\beta)(m_1^2\beta+m_2^2\alpha-\alpha\beta
s)}{6144\pi^6\alpha\beta}\left(\frac{3m_1^2\beta+3m_2^2\alpha-5\alpha\beta
s}{\beta}-\frac{3m_1^2\beta+3m_2^2\alpha-9\alpha\beta
s}{\alpha}\right), \non
\rho_{2a}^{\qGqb}(s)&=&\qGqa\dab\left[\frac{m_2(2m_1^2\beta+2m_2^2\alpha-3\alpha\beta
s)}{32\pi^4\beta}+\frac{m_1(m_1^2\beta+m_2^2\alpha-2\alpha\beta
s)}{32\pi^4\alpha}\right], \non
\rho_{2b}^{\qGqb}(s)&=&\qGqa\dab\Bigg[\frac{m_1(1-\alpha-\beta)(m_1^2\beta+m_2^2\alpha-2\alpha\beta
s)}{64\pi^4\alpha^2}-\frac{m_qm_1m_2}{64\pi^4\beta}\Bigg], \non
\rho_{2c}^{\qGqb}(s)&=&-\frac{m_qm_1m_2\qGqa}{16\pi^4}\left[\left(1+\frac{m_1^2-m_2^2}{s}\right)^2-\frac{4m_1^2}{s}\right]^{1/2},
\non
\rho_2^{\qq^2}(s)&=&\frac{\qq^2}{24\pi^2}\left[\left(1+\frac{m_1^2-m_2^2}{s}\right)^2-\frac{4m_1^2}{s}\right]^{1/2}
\Bigg\{4m_1m_2-\frac{m_qm_1(m_1^2-m_2^2-s)+2m_qm_2(m_1^2-m_2^2+s)}{s}
\non&&+\frac{2m_qm_1}{s[(m_1^2-m_2^2-s)^2-4m_2^2s]}\left[m_1^6-m_2^6+m_2^4s-m_1^4(3m_2^2+2s)+m_1^2(3m_2^4+m_2^2s+s^2)\right]
\non&&+\frac{2m_qm_2}{s[(m_1^2-m_2^2-s)^2-4m_2^2s]}\left[m_1^6-m_1^4(3m_2^2+s)+m_1^2(3m_2^4-m_2^2s)-(m_2^3-m_2s)^2\right]\Bigg\},
\non
\rho_{2a}^{\qq\qGqb}(s)&=&\frac{\qq\qGqb}{12\pi^2}\int_0^1d\alpha\frac{m_1m_2^3}{\alpha^2}\delta'\left[s-\frac{m_1^2\alpha+m_2^2(1-\alpha)}{\alpha(1-\alpha)}\right],
\\
\rho_{2b}^{\qq\qGqb}(s)&=&\frac{\qq\qGqb}{48\pi^2}\int_0^1d\alpha\frac{m_1m_2}{\alpha}\delta\left[s-\frac{m_1^2\alpha+m_2^2(1-\alpha)}{\alpha(1-\alpha)}\right].
\end{eqnarray}
}

For the interpolating current $J_4$ with $J^P=1^+$:
\begin{eqnarray}
\nonumber\rho_4^{pert}(s)&=&\frac{1}{2}\rho_2^{pert}(s),
\rho_4^{\qq}(s)=\frac{1}{2}\rho_2^{\qq}(s),
\rho_{4a}^{\GGa}(s)=\frac{1}{2}\rho_{2a}^{\GGa}(s),
\rho_{4b}^{\GGa}(s)=-\rho_{2b}^{\GGa}(s), \non
\rho_{4a}^{\qGqb}(s)&=&\frac{1}{2}\rho_{2a}^{\qGqb}(s),
\rho_{4b}^{\qGqb}(s)=-\rho_{2b}^{\qGqb}(s),
\rho_{4c}^{\qGqb}(s)=\frac{1}{2}\rho_{2c}^{\qGqb}(s),
\rho_4^{\qq^2}(s)=\frac{1}{2}\rho_2^{\qq^2}(s),
\\ \rho_{4a}^{\qq\qGqb}(s)&=&\frac{1}{2}\rho_{2a}^{\qq\qGqb}(s), \rho_{4b}^{\qq\qGqb}(s)=-\rho_{2b}^{\qq\qGqb}(s). \label{eq2}
\end{eqnarray}

For the mixed interpolating currents $J^m$ and $J^m_{\mu}$, we just
calculate the mixed parts $\langle
0|T[J_1J_2^{\dag}]|0\rangle+\langle 0|T[J_2J_1^{\dag}]|0\rangle$ and
$\langle 0|T[J_{1\mu}J_{2\nu}^{\dag}]|0\rangle+\langle
0|T[J_{2\mu}J_{1\nu}^{\dag}]|0\rangle$ in the correlation functions.
The mixed part of the spectral density for $J^m$ with $J^P=0^+$ is
{\allowdisplaybreaks
\begin{eqnarray}
\nonumber\rho_m^{pert}(s)&=&0, \non \rho_m^{\qq}(s)&=&0, \non
\rho_{m}^{\GGa}(s)&=&\GGb\dab\frac{5m_1m_2(m_1^2\beta+m_2^2\alpha-\alpha\beta
s)}{1024\pi^6\alpha\beta}
\left[\frac{1-\beta}{\alpha}+\frac{1-\alpha}{\beta}-\frac{(1-\alpha-\beta)^2}{2\alpha\beta}-3\right],
\non \rho_{m}^{\qGqb}(s)&=&\qGqa\dab
\non&&\frac{5(2m_1^2\beta+2m_2^2\alpha-3\alpha\beta
s)}{128\pi^4}\left[\frac{m_1(1-\beta)}{\alpha^2}+\frac{m_2(1-\alpha)}{\beta^2}
-\frac{2m_1-m_q}{\alpha}-\frac{2m_2-m_q}{\beta}\right], \\
\rho_m^{\qq^2}(s)&=&0, \non
\rho_{m}^{\qq\qGqb}(s)&=&\frac{5\qq\qGqb}{48\pi^2}\int_0^1d\alpha\left\{\frac{m_1^2\alpha+m_2^2(1-\alpha)}{2\alpha(1-\alpha)}
\delta\left[s-\frac{m_1^2\alpha+m_2^2(1-\alpha)}{\alpha(1-\alpha)}\right]+H\left[s-\frac{m_1^2\alpha+m_2^2(1-\alpha)}{\alpha(1-\alpha)}\right]\right\}.
\end{eqnarray}
}

For the mixed interpolating current $J^m_{\mu}$ with $J^P=1^+$:
{\allowdisplaybreaks
\begin{eqnarray}
\nonumber\rho_m^{pert}(s)&=&0, \non \rho_m^{\qq}(s)&=&0, \non
\rho_{m}^{\GGa}(s)&=&\GGb\dab\frac{5m_1m_2(m_1^2\beta+m_2^2\alpha-\alpha\beta
s)}{36864\pi^6\alpha\beta}
\non&&\left[\frac{(1-\alpha-\beta)^2(5+\alpha+\beta)}{\alpha\beta}-\frac{3(1-\alpha-\beta)(3+\alpha+\beta)(\alpha+\beta)}{\alpha\beta}
+6(1+\alpha+\beta)\right], \non \rho_{m}^{\qGqb}(s)&=&\qGqa\dab
\non&&\frac{5(3m_1^2\beta+3m_2^2\alpha-5\alpha\beta
s)}{768\pi^4}\left[\left(\frac{m_1-m_q}{\alpha}+\frac{m_2-m_q}{\beta}\right)-\left(\frac{m_1}{\alpha^2}+\frac{m_2}{\beta^2}\right)(1-\alpha-\beta)\right],
\\ \rho_m^{\qq^2}(s)&=&0, \non
\rho_{m}^{\qq\qGqb}(s)&=&-\frac{5\qq\qGqb}{576\pi^2}\int_0^1d\alpha\left\{\frac{2\left[m_1^2\alpha+m_2^2(1-\alpha)\right]}{\alpha(1-\alpha)}
\delta\left[s-\frac{m_1^2\alpha+m_2^2(1-\alpha)}{\alpha(1-\alpha)}\right]+3*H\left[s-\frac{m_1^2\alpha+m_2^2(1-\alpha)}{\alpha(1-\alpha)}\right]\right\}.
\end{eqnarray}
}

\end{document}